# Predictions on the alpha decay half lives of Superheavy nuclei with Z = 113 in the range 255 ≤ A ≤ 314

K. P. Santhosh*, A. Augustine, C. Nithya and B. Priyanka

*School of Pure and Applied Physics, Kannur University, Swami Anandatheertha Campus, Payyanur 670327, Kerala, India*

**Abstract**

An intense study of the alpha decay properties of the isotopes of superheavy element Z=113 have been performed within the Coulomb and proximity potential model for deformed nuclei (CPPMDN) within the wide range 255 ≤ A ≤ 314. The predicted alpha decay half lives of $^{278}113$ and $^{282}113$ and the alpha half lives of their decay products are in good agreement with the experimental data. 6α chains and 4α chains predicted respectively for $^{278}113$ and $^{282}113$ are in agreement with the experimental observation. Our study shows that the isotopes in the mass range 278 ≤ A ≤ 286 will survive fission and can be synthesized and detected in the laboratory via alpha decay. In our study, we have predicted 6α chains from $^{279}113$, 4α chains from $^{286}113$, 3α chains from $^{280,281,283}113$, 2α chains from $^{284}113$ and 1α chain from $^{285}113$. We hope that these predictions will be a guideline for future experimental investigations.

*email: drkpsanthosh@gmail.com

## 1. Introduction

Superheavy nuclei (SHN) and their decay studies is one of the fast developing fields in nuclear physics. Significant theoretical and experimental investigations have been made in the region of superheavy nuclei in predicting the existence of magic island or island of stability [1-5]. Recently the isotopes of many superheavy elements have been synthesized successfully through hot fusion reactions [6], performed at JINR, FLNR (Dubna) and cold fusion reactions [7], performed at GSI (Darmstadt, Germany). The concept of cold fusion was proposed in 1970s and realized experimentally in 1980s. In cold fusion reaction the heaviest superheavy element so far synthesized is the isotope of Z =113 ($^{278}113$) by Morita et al. in 2004 [8] and the synthesis of $^{278}113$ is confirmed in 2012 [9]. This has been recently accepted by IUPAC and IUPAP [10].

One of the fundamental questions in nuclear physics is about the number of possible elements that can be found in nature or that can be produced in the laboratory. Two different approaches, that is, the hot fusion approach and the cold fusion approach were used recently to extend the periodic table. The elements with Z = 107-112 were synthesized using the cold fusion approach. Attempts to synthesize heavier elements via cold fusion were unsuccessful because of the limited beam time of accelerators for superheavy nuclei beyond Z = 112. First attempt to synthesize the element Z=113 by cold fusion reaction was done at velocity filter SHIP at GSI, Darmstadt. Three experimental runs were performed altogether in the period 1998-2003, without observing a single decay chain starting from an isotope of the element Z=113. Morita et al. [8] started the experiments to synthesize the element Z=113 at the gas filled separator GARIS, RIKEN, using $^{209}$Bi ($^{70}$Zn, n) reaction, in September 5, 2003 and the first decay chain of the element had been observed in 2004, which was interpreted to start from $^{278}113$. In 2007, Oganessian et al. [11] were successful in producing the element $^{282}113$ by hot fusion reaction, using $^{48}$Ca projectile on actinide target $^{237}$Np, at the Flerov Laboratory of Nuclear Reaction (FLNR) of Joint Institute of Nuclear Research (JINR), Russia and its alpha chains has been observed.

The superheavy nuclei decay mainly by the emission of alpha particles followed by subsequent spontaneous fission. Studies on the characteristic alpha chains will help in the identification of new nuclides. The phenomenon of alpha decay was discovered by Rutherford [12, 13] in 1899 and was first described by Gamow [14] in 1928 using the idea of quantum tunneling through the potential barrier. Extensive experimental [15-23] and theoretical works [24-37] have been performed in order to understand the formation of superheavy nuclei and their alpha decay half lives. The formation of superheavy nuclei can be successfully explained by dinuclear system (DNS) concept, in which the fusion process is assumed as a transfer of nucleons from the light nucleus to the heavy one [38-42]. Using DNS model Adamain et al. [43] presented the calculations on the production cross sections for the heaviest nuclei and suggested the reaction $Zn^{68}$+$Bi^{209}$ for the synthesis of the isotope $^{279}113$. Based on DNS model, production cross section of superheavy nuclei Z = 112-116 in $^{48}Ca$ induced reaction is studied by Feng et al. [44]. The studies on the synthesis of superheavy nuclei with Z = 119 and 120 [45]; and 118 [46] was done by Wang et al. within the dinuclear system with dynamical potential energy surface model (DNS-DyPES model).

A number of works have been performed to study the properties of odd Z superheavy nuclei [47-55]. The structure of the nuclide with Z = 105 and its alpha decay chain was studied systematically by Long et al. [52] within the relativistic mean field approach (RMF) in 2002. Within the density dependent cluster model, calculations on the alpha decay half lives of the heaviest odd Z elements $115 \rightarrow 113 \rightarrow Rg$ was done by Ren et al [31]. Using macroscopic-microscopic model Peng et al. [51] studied alpha decay of 323 nuclei with Z ≥ 82 which includes the isotopes of odd Z elements, Z=107-115.

Theoretical studies on the alpha decay properties of Z = 113 have been done by Tai et al. [56] within the frame work of density dependent cluster model (DDCM) with renormalized RM3Y nucleon – nucleon interactions (RM3Y) [57] and by Dong et al. [58] using cluster model and generalized liquid drop model (GLDM).

The intention of our present work is to compare the alpha decay half lives and spontaneous fission half lives of various isotopes of the superheavy element Z = 113 and to predict the decay modes, using the Coulomb and proximity potential model for deformed nuclei (CPPMDN) [59], which is an extension of Coulomb and proximity potential model (CPPM), proposed by Santhosh et al. [60]. Our previous works on the decay properties of heavy and superheavy nuclei [61-67] has revealed the success and applicability of CPPMDN formalism in predicting the decay half lives. The agreement between experimental and theoretical results is also discussed in detail.

The overview of the paper is as follows. In Sec 2, we briefly describe the features of Coulomb and proximity potential model for deformed nuclei. The results and discussion on the alpha decay properties of various isotopes of the superheavy element Z = 113 are presented in Sec 3 and a brief summary of the entire work is given in the last section.

## 2. The Coulomb and proximity potential model for deformed nuclei (CPPMDN)

The interacting potential between two nuclei in CPPMDN is taken as the sum of deformed Coulomb potential, deformed two term proximity potential and centrifugal potential, for both the touching configuration and for the separated fragments. For the pre-scission (overlap) region, simple power law interpolation as done by Shi and Swiatecki [68] has been used. It was observed [60] that the inclusion of the proximity potential reduces the height of the potential barrier, which agrees with the experimental result.

Shi and Swiatecki [68] were the first to use the proximity potential in an empirical manner and later on, several theoretical groups [69-71] have used the proximity potential, quite extensively for various studies including the fusion excitation function. The contribution of both the internal and the external part of the barrier has been considered, in the present model, for the penetrability calculation and the assault frequency, $\nu$ is calculated for each parent-cluster combination which is associated with the vibration energy. However, for even A parents and for odd A parents, Shi and Swiatecki [72] get $v$ empirically, unrealistic values as $10^{22}$ and $10^{20}$, respectively.

The interacting potential barrier for two spherical nuclei is given by

$$V = \frac{Z_1 Z_2 e^2}{r} + V_p(z) + \frac{\hbar^2 \ell(\ell+1)}{2\mu r^2}, \quad \text{for } z > 0 \tag{1}$$

Here $Z_1$ and $Z_2$ are the atomic numbers of the daughter and emitted cluster, '$r$' is the distance between fragment centres, '$z$' is the distance between the near surfaces of the fragments, $\ell$ represents the angular momentum and $\mu$ the reduced mass. $V_P$ is the proximity potential given by Blocki *et al.*, [73, 74] as,

$$V_p(z) = 4\pi\gamma b\left[\frac{C_1 C_2}{(C_1 + C_2)}\right]\Phi\left(\frac{z}{b}\right) \tag{2}$$

with the nuclear surface tension coefficient,

$$\gamma = 0.9517\,[1 - 1.7826\,(N - Z)^2 / A^2]\ \text{MeV/fm}^2 \tag{3}$$

Here $N$, $Z$ and $A$ represent the neutron, proton and mass number of the parent and $\Phi$ represents the universal proximity potential [74] given as

$$\Phi(\varepsilon) = -4.41 e^{-\varepsilon/0.7176}\ , \text{for } \varepsilon \geq 1.9475 \tag{4}$$

$$\Phi(\varepsilon) = -1.7817 + 0.9270\varepsilon + 0.0169\varepsilon^2 - 0.05148\varepsilon^3\ , \text{for } 0 \leq \varepsilon \leq 1.9475 \tag{5}$$

With $\varepsilon = z/b$, where the width (diffuseness) of the nuclear surface $b \approx 1$ fermi and the Süsmann central radii $C_i$ of the fragments are related to the sharp radii $R_i$ as

$$C_i = R_i - \left(\frac{b^2}{R_i}\right) \text{ fm} \tag{6}$$

For $R_i$, we use semi-empirical formula in terms of mass number $A_i$ as [73]

$$R_i = 1.28 A_i^{1/3} - 0.76 + 0.8 A_i^{-1/3} \text{ fm} \tag{7}$$

The potential for the internal part (overlap region) of the barrier is given as,

$$V = a_0 (L - L_0)^n\ , \text{for } z < 0 \tag{8}$$

where $L = z + 2C_1 + 2C_2$ fm and $L_0 = 2C$ fm, the diameter of the parent nuclei. The constants $a_0$ and $n$ are determined by the smooth matching of the two potentials at the touching point.

Using the one dimensional Wentzel-Kramers-Brillouin approximation, the barrier penetrability $P$ is given as

$$P = \exp\left\{-\frac{2}{\hbar}\int_a^b \sqrt{2\mu(V - Q)}\,dz\right\} \tag{9}$$

Here the mass parameter is replaced by $\mu = mA_1 A_2 / A$, where $m$ is the nucleon mass and $A_1$, $A_2$ are the mass numbers of daughter and emitted cluster respectively. The turning points "$a$" and "$b$" are determined from the equation, $V(a) = V(b) = Q$. The above integral can be evaluated numerically or analytically, and the half life time is given by

$$T_{1/2} = \left(\frac{\ln 2}{\lambda}\right) = \left(\frac{\ln 2}{\nu P}\right) \tag{10}$$

where, $\nu = \left(\frac{\omega}{2\pi}\right) = \left(\frac{2E_v}{h}\right)$ represent the number of assaults on the barrier per second and $\lambda$ the decay constant. $E_v$, the empirical vibration energy is given as [75]

$$E_v = Q\left\{0.056 + 0.039 \exp\left[\frac{(4 - A_2)}{2.5}\right]\right\}, \quad \text{for } A_2 \geq 4 \tag{11}$$

Classically, the $\alpha$ particle is assumed to move back and forth in the nucleus and the usual way of determining the assault frequency is through the expression given by $\nu = velocity/(2R)$, where $R$ is the radius of the parent nuclei. As the alpha particle has wave properties, a quantum mechanical treatment is more accurate. Thus, assuming that the alpha particle vibrates in a harmonic oscillator potential with a frequency $\omega$, which depends on the vibration energy $E_v$, we can identify this frequency as the assault frequency $v$ given in equations (10) and (11).

The Coulomb interaction between the two deformed and oriented nuclei with higher multipole deformation included [76, 77] is taken from Ref. [78] and is given as,

$$V_C = \frac{Z_1 Z_2 e^2}{r} + 3Z_1 Z_2 e^2 \sum_{\lambda,i=1,2} \frac{1}{2\lambda+1} \frac{R_{0i}^{\lambda}}{r^{\lambda+1}} Y_{\lambda}^{(0)}(\alpha_i) \left[ \beta_{\lambda i} + \frac{4}{7} \beta_{\lambda i}^2 Y_{\lambda}^{(0)}(\alpha_i) \delta_{\lambda,2} \right] \tag{12}$$

with

$$R_i(\alpha_i) = R_{0i} \left[ 1 + \sum_{\lambda} \beta_{\lambda i} Y_{\lambda}^{(0)}(\alpha_i) \right] \tag{13}$$

where $R_{0i} = 1.28 A_i^{1/3} - 0.76 + 0.8 A_i^{-1/3}$. Here $\alpha_i$ is the angle between the radius vector and symmetry axis of the $i^{\text{th}}$ nuclei (see Fig.1 of Ref [76]) and it is to be noted that the quadrupole interaction term proportional to $\beta_{21}\beta_{22}$, is neglected because of its short-range character.

The proximity potential and the double folding potential can be considered as the two variants of the nuclear interaction [79, 38]. In the description of interaction between two fragments, the latter is found to be more effective. The proximity potential of Blocki *et al.*, [73, 74], which describes the interaction between two pure spherically symmetric fragments, has one term based on the first approximation of the folding procedure and the two-term proximity potential of Baltz *et al.*, (equation (11) of [80]) includes the second component as the second approximation of the more accurate folding procedure. The authors have shown that the two-term proximity potential is in excellent agreement with the folding model for heavy ion reaction, not only in shape but also in absolute magnitude (see figure 3 of [80]). The two-term proximity potential for interaction between a deformed and spherical nucleus is given by Baltz *et al.*, [80] as

$$V_{P2}(R,\theta) = 2\pi \left[ \frac{R_1(\alpha) R_C}{R_1(\alpha) + R_C + S} \right]^{1/2} \left[ \frac{R_2(\alpha) R_C}{R_2(\alpha) + R_C + S} \right]^{1/2} \times \left[ \left[ \varepsilon_0(S) + \frac{R_1(\alpha) + R_C}{2R_1(\alpha) R_C} \varepsilon_1(S) \right] \left[ \varepsilon_0(S) + \frac{R_2(\alpha) + R_C}{2R_2(\alpha) R_C} \varepsilon_1(S) \right] \right]^{1/2} \tag{14}$$

where $\theta$ is the angle between the symmetry axis of the deformed nuclei and the line joining the centers of the two interacting nuclei, and $\alpha$ corresponds to the angle between the radius vector and symmetry axis of the nuclei (see Fig. 5 of Ref [80]). $R_1(\alpha)$ and $R_2(\alpha)$ are the principal radii of curvature of the daughter nuclei, $R_C$ is the radius of the spherical cluster, $S$ is the distance between the surfaces along the straight line connecting the fragments, and $\varepsilon_0(S)$ and $\varepsilon_1(S)$ are the one dimensional slab-on-slab function.

The barrier penetrability of α particle in a deformed nucleus is different in different directions. The averaging of penetrability over different directions is done using the equation

$$P = \frac{1}{2} \int_0^{\pi} P(Q,\theta,\ell) \sin(\theta) d\theta \tag{15}$$

where $P(Q,\theta,\ell)$ is the penetrability of α particle in the direction $\theta$ from the symmetry axis for axially symmetric deformed nuclei.

## 3. Results and discussion

Studies on the decay properties of superheavy nuclei provide information on their existence and stability in nature. The investigations on the half lives of different radioactive decay play a significant role in determining the properties of superheavy nuclei. The dominant decay modes of superheavy nuclei involve alpha decay and spontaneous fission. Several theoretical models are available for calculating the alpha decay half lives as well as spontaneous fission half lives. It is seen that those nuclei with small alpha decay half lives than the spontaneous fission half lives survive fission and thus can be detected in laboratories via alpha decay.

### 3.1 Alpha Decay Half lives

In the present study the alpha half lives of the isotopes of SHN with Z = 113 have been studied within the range 255≤ A ≤314 using CPPMDN and the present values are then compared with those calculated by means of CPPM [60], Viola-Seaborg semiempirical (VSS) relationship [81], The Universal (UNIV) curve of Poenaru et al. [82, 83] and the analytical formula of Royer [84].

The alpha decay is characterised by the energy release $Q_\alpha$ and the corresponding life time $T_\alpha$. In alpha transitions, Q value is the energy released between the ground state energy levels of the parent nuclei and ground state energy levels of the daughter nuclei and is given as,

$$Q = \Delta M_p - (\Delta M_\alpha + \Delta M_d) + k(Z_p^\varepsilon - Z_d^\varepsilon) \tag{16}$$

which is positive for a given decay. Here $\Delta M_p$, $\Delta M_d$, $\Delta M_\alpha$ are the mass excess of the parent, daughter and alpha particle respectively. In order to calculate the Q value, the mass excesses are taken from Ref [85, 86]. The electron screening correction [87] have been included by the term $k(Z_p^\varepsilon - Z_d^\varepsilon)$, where $k$ = 8.7eV , $\varepsilon$ =2.517 for $Z \geq 60$ and $k$ = 13.6eV, $\varepsilon$ = 2.408 for $Z < 60$. The quadrupole ($\beta_2$) and hexadecapole ($\beta_4$) deformation values of the parent and daughter nuclei have been used for the calculation of alpha half lives and the deformation values taken from Ref. [88] are used for the calculation.

The well known Viola-Seaborg semi-empirical Relationship (VSS) formula for calculating the alpha decay half lives is given by,

$$\log_{10}(T_{1/2}) = (aZ + b)Q^{-1/2} + cZ + d + h_{\log} \tag{17}$$

Here the half life is in seconds and the Q value is in MeV. Z is the atomic number of the parent nucleus, a, b, c, d, $h_{\log}$ are adjustable parameters. The quantity $h_{\log}$ gives the hindrance of alpha decay for the nuclei with odd proton and odd neutron numbers [81]. Instead of using the original set of constants given by Viola and Seaborg [81], more recent values determined by Sobiczewski et al. [89] has been used here. The constants are a = 1.66175, b = -8.5166, c = -0.20228, d = -33.9069 and

$$h_{\log} = \begin{cases} 0, & for \quad Z = even \quad N = even \\ 0.772, & for \quad Z = odd \quad N = even \\ 1.066, & for \quad Z = even \quad N = odd \\ 1.114, & for \quad Z = odd \quad N = odd \end{cases} \tag{18}$$

For calculating the decay half lives several simple and effective relationships are available, which are obtained by fitting experimental data. Among them one of the important relationship is the UNIV curves [90-93], derived by extending a fission theory to larger mass asymmetry. Based on the quantum mechanical tunnelling process, the relationship [94, 95] of the disintegration constant λ, valid in both fission like and α-like theories, and the partial decay half life T of the parent nucleus is given as,

$$\lambda = \ln 2 / T = \nu S P_S \tag{19}$$

Here ν, S and $P_S$ are three model dependent quantities. ν is the frequency of assaults on the barrier per second, S is the pre-formation probability of the cluster at the nuclear surface (equal to the probability of the internal part of the barrier in a fission theory [90, 91]), and $P_S$ is the quantum penetrability of the external potential barrier.

By using the decimal logarithm equation (18) becomes,

$$\log_{10} T(s) = -\log_{10} P - \log_{10} S + [\log_{10}(\ln 2) - \log_{10} \nu] \quad (20)$$

To derive the universal formula, the basic assumptions were that ν = constant and S depends only on the mass number of emitted particle $A_e$ [91, 94]. It was shown by a macroscopic calculation of pre-formation probability [96] of many clusters from $^8$Be to $^{46}$Ar that, $A_e$ depends only upon the size of the cluster. Using a fit with experimental data for α decay, the corresponding numerical values [91] had been obtained: $s_\alpha$ = 0.0143153, $\nu = 10^{22.01}$ s$^{-1}$. The additive constant for even-even nuclei is given as,

$$c_{ee} = [-\log_{10} \nu + \log_{10}(\ln 2)] = -22.16917 \quad (21)$$

And the decimal logarithm of the pre-formation factor is

$$\log_{10} S = -0.598(A_e - 1) \quad (22)$$

The penetrability of an external Coulomb barrier, having the separation at the touching configuration $R_a = R = R_d + R_e$ as the first turning point, and the second one defined by $e^2 Z_d Z_e / R_b = Q$ may be obtained analytically as,

$$-\log_{10} P_S = 0.22873(\mu_A Z_d Z_e R_b)^{1/2} \times [\arccos\sqrt{r} - \sqrt{r(1-r)}] \quad (23)$$

where $r = R_t / R_b$ fm, $R_t = 1.2249(A_d^{1/3} + A_e^{1/3})$ fm and $R_b = 1.43998 Z_d Z_e / Q$ fm.

To calculate the released energy Q, the liquid drop model radius constant $r_0$ = 1.2249 fm and the mass tables [85, 86] are used.

Geiger and Nuttal [97] formulated the earliest law for the alpha decay half lives. Several expressions [81, 89, 98, 99] were advanced subsequently. Royer [84] formulated the following formula by a fitting procedure applied on a set of 373 alpha emitters with a RMS deviation of 0.42

$$\log_{10}[T_{1/2}(s)] = -26.06 - 1.114 A^{1/6}\sqrt{Z} + \frac{1.5837Z}{\sqrt{Q_\alpha}} \quad (24)$$

Here A and Z are the mass and charge numbers of the parent nuclei and $Q_\alpha$ is the energy released during the reaction.

The following relation corresponds to a subset of 86 odd-even nuclei and a RMS deviation of 0.36

$$\log_{10}[T_{1/2}(s)] = -25.68 - 1.1423 A^{1/6}\sqrt{Z} + \frac{1.592Z}{\sqrt{Q_\alpha}} \quad (25)$$

For a subset of 50 odd-odd nuclei the RMS deviation was found to be 0.35 and the formula is given by,

$$\log_{10}[T_{1/2}(s)] = -29.48 - 1.113 A^{1/6}\sqrt{Z} + \frac{1.6971Z}{\sqrt{Q_\alpha}} \quad (26)$$

### 3.2 Spontaneous fission half lives

The spontaneous fission (SF) half-lives of various nuclei can be calculated by using the semi-empirical relation given by Xu et al [100]. The equation was originally made to fit the even-even nuclei and is given as,

$$T_{1/2} = \exp\left\{2\pi\left[C_0 + C_1 A + C_2 Z^2 + C_3 Z^4 + C_4 (N-Z)^2 - (0.13323\frac{Z^2}{A^{1/3}} - 11.64)\right]\right\} \quad (27)$$

Here the constants $C_0$ = -195.09227, $C_1$ = 3.10156, $C_2$ = -0.04386, $C_3$ = 1.4030 x 10$^{-6}$ and $C_4$ = -0.03199. In the present work we have considered only the odd mass (i.e odd-even and odd-odd nuclei) nuclei. So instead of taking spontaneous fission half lives directly, we have taken the average of spontaneous fission half lives of corresponding neighboring even-even nuclei. In the case of odd-even nuclei, we took the $T_{sf}^{av}$ of two neighboring even-even nuclei and while dealing with odd-odd nuclei, the $T_{sf}^{av}$ of four neighboring even-even nuclei was taken.

Attempts to synthesize the superheavy element Z=113 started as early as 2003. The isotope $^{278}113$ was produced through $^{207}Np+^{70}Zn$ reaction with six consecutive alpha chains [9]. The $^{282}113$ nuclide was synthesized through $^{237}Np+^{48}Ca$ fusion reaction and consequently its alpha decay chains were observed [101]. Various isotopes of the element Z=113 namely $^{283}113$ and $^{284}113$ have been observed in the decay chains of isotopes of Z=115 and the isotopes $^{285}113$, $^{286}113$ have been observed in the decay chains of isotopes of Z=117 [102]. In the present paper we compare the alpha decay half lives and spontaneous fission half lives of various isotopes of Z=113 in order to find the mode of decay of these nuclides, concentrating mainly on the recently synthesized $^{278,\ 282}113$ isotopes and then theses were compared with experimental data. The comparison of spontaneous fission half lives and alpha decay half lives calculated within our model and the predictions on the decay chains are given in Table 1. The comparison of the present values with other theoretical models is also shown.

In Table 1 the first column denotes the parent and daughter nuclei. Column 2 gives experimental Q values of these isotopes taken from Ref [9, 101]. The spontaneous fission half lives of the isotopes under study evaluated using the phenomenological formula of Xu et al. is given in column 3. Experimental alpha decay half lives obtained from [9, 101] are arranged in column 4. Column 5 shows the alpha decay half lives of these isotopes calculated using CPPMDN formalism. The alpha half life calculations using CPPM are given in column 6. In CPPMDN the nucleus-nucleus interaction potential is calculated using equation (14), while in CPPM (spherical case) the potential is calculated using equation (2). On comparing the alpha decay half lives calculated within both these formalisms we can see that the alpha half lives decrease with the inclusion of deformation values. Within our fission model the pre-formation probability, $S$ [103, 104] can be calculated as the penetrability of the internal part (overlap region) of the barrier given as

$$S = \exp(-K) \tag{28}$$

Where

$$K = \frac{2}{\hbar}\int_{a}^{0}\sqrt{2\mu(V-Q)}\,dz \tag{29}$$

here, $a$ is the inner turning point and is defined as $V(a)=Q$ and $z=0$ represents the touching configuration. The VSS, analytical formula of Royer and UNIV have also been used for determining the alpha decay half lives and are given in columns 7, 8 and 9 respectively. The last column represents the mode of decay of isotopes under study. From the table, it is clear that, by comparing the SF half lives with the alpha decay half lives we can predict a 6α chains from the isotope $^{278}113$, which agrees well with the experimental observation. Experimentally it was shown that after the 6$^{th}$α chain, the isotope $^{254}Md$ shows electron capture ($b_{\varepsilon}$ = 100%) [105] and thereafter the daughter isotope $^{254}Fm$ will undergo alpha decay. The same result has been predicted within CPPMDN. In the case of $^{282}113$, it can be clearly seen that the alpha decay half lives computed within CPPMDN closely agrees with the experimental values. By comparing the SF half lives calculated using the semi-empirical relation given by Xu et al. with the alpha decay half lives we can predict α chains from the isotope, but for a more accurate prediction on the decay mode, we have used the values given by Smolanczuk et al. [106, 107], in which the spontaneous fission half lives of even-even nuclei with Z=104-114 has analyzed in a multidimensional deformation space, in a dynamical approach without any adjustable parameters. Using these values, the average spontaneous fission half lives were calculated, and on comparing the alpha decay half lives with the corresponding spontaneous fission half lives we can predict 4α chains for the isotope $^{282}113$, which matches very well with the experimental result. So by using our formalism, even though there is a one order difference in alpha decay half lives for some of the isotopes under study, the predictions on the alpha decay half lives and decay modes of the experimentally synthesized $^{278}113$ and $^{282}113$ go hand in hand with the experimental results. Thus we extended our work to predict the alpha decay half lives and mode of decay of 58 more isotopes of Z = 113, ranging from 255 ≤ A ≤ 314.

Figures 1-15 represents the entire work. We have plotted $\log_{10}T_{1/2}$ against the mass number of the parent nuclei. All the calculations done within the various theoretical models are shown. It is to be noted that the decay half lives evaluated by using VSS formula, UNIV and the analytical formula of Royer match well with our theoretical calculations.

Figure 1-3 shows the plot of $\log_{10}T_{1/2}$ versus mass number for the parent nuclei $^{255\text{-}266}113$ and their decay products. By comparing the alpha decay half lives with the corresponding spontaneous fission half lives, it can be clearly seen that none of these isotopes will survive fission. In figure 4, the plots of isotopes $^{267\text{-}270}113$ are shown. We can see that the isotopes $^{267\text{-}269}113$ will not survive fission, whereas the isotope $^{270}113$ will survive fission and shows full alpha chain within CPPMDN. Figures 5 and 6, shows the plot for the isotopes $^{271\text{-}278}113$, which include the experimentally synthesized SHN $^{278}113$. It is clear from the figure that all these isotopes will survive fission and show full alpha chain within CPPMDN. But in the case of $^{278}113$, it was seen that after the 6$^{th}$ chain the daughter isotope, $^{254}$Md, undergoes electron capture**.** Even though the isotopes $^{270\text{-}277}113$ decay by emitting alpha particles, they are hard to detect in laboratory because of their small decay times (for e.g., $T_{1/2}^{\alpha}$= 3.059x10$^{-8}$s for $^{270}113$ and $T_{1/2}^{\alpha}$ = 1.320x10$^{-8}$s for $^{271}113$). The calculations done for the experimentally synthesized $^{278}113$ is shown in figure 6(d). Experimental alpha decay values have been represented as scattered points in the figure. Plot for the isotopes $^{279\text{-}282}113$ are shown in figure 7. It is seen that the isotope $^{279}113$ shows full alpha chain within CPPMDN. But after the 6$^{th}$ alpha chain the isotope $^{255}$Md shows electron capture ($b_{\varepsilon}$ = 92%) [105] and thereafter the daughter isotope $^{255}$Fm will undergo alpha decay. The isotopes $^{280,\,281}113$ will survive fission and shows 3α chains by comparing the alpha decay half lives with the spontaneous fission half lives of Xu et al. In the case of $^{282}113$, we got 4α chains as mentioned earlier. The half lives for these isotopes are in millisecond range (in the case of $^{280}113$ $T_{1/2}^{\alpha}$ = 7.131 x 10$^{-4}$s, for $^{281}113$ $T_{1/2}^{\alpha}$ = 1.635 x 10$^{-3}$s and for $^{282}113$ $T_{1/2}^{\alpha}$ = 4.873 x 10$^{-3}$s) and hence can be synthesized and detected via alpha decay in laboratory. Figure 7(d) represents the plot of experimentally synthesized $^{282}113$. The scattered points in the figure represent experimental alpha decay values. The average spontaneous fission values given by Xu et al. and Smolanczuk et al. are also shown. Figure 8 depicts the decay properties of isotopes $^{283\text{-}286}113$. From the figure it is clear that the isotopes $^{283\text{-}285}113$ will survive fission and 3α, 2α and 1α chains can be predicted respectively from the isotopes $^{283}113$, $^{284}113$ and $^{285}113$. These isotopes can be detected in laboratory through alpha decay because of their longer alpha half lives. It is to be noted that our theoretical predictions on the alpha decay half lives and decay modes of the nuclei $^{283}113$ and $^{284}113$ matches well with the experimental values of these isotopes, which were obtained as the decay products of $^{288}115$ and $^{287}115$ respectively [108], and the comparison between experimental and theoretical results are given in detail in Table 1 of our previous work Ref [61]. Similarly the isotopes $^{285}113$ and $^{286}113$ were observed as the decay products of the isotopes $^{293}117$ and $^{294}117$ isotopes respectively [23]. It is seen that the alpha decay half lives calculated within CPPMDN is in good agreement with the experimental results. In the case of $^{286}113$, for a better matching with experimental decay modes, we have adopted the spontaneous fission values given in [106, 107]. 4α chains can be predicted from the isotope by comparing the alpha decay half lives with the spontaneous fission half lives and it is evident that the predictions on the decay modes of the isotope is same as the experimental results. The comparison between experimental and theoretical values of alpha decay half lives and decay modes are given in Table 1 of Ref [63]. Figures 9-15 represents the plots for the isotopes 287 ≤A ≤ 314. We can see that none of these isotopes will survive fission and it is hard to observe them in laboratories. Thus the nuclei within the range 278 ≤ A ≤ 286 were found to have relatively long alpha decay half-lives and can be detected in laboratory. These predictions are included in Table II and Table III. Table II shows the comparison of the spontaneous fission half lives with the alpha decay half lives for the nuclei $^{279-281,283,284}$113 and Table III shows the same for $^{285,286}$113 nuclei. We have also included the predictions on the decay modes of these isotopes within CPPMDN, which will be helpful in future experimental investigations. The pictorial representation of alpha decay chains of predicted isotopes are shown in figure 16.

We hope that our present study, which predicts the mode of decay of various isotopes of Z = 113 within a wide range 255 ≤ A ≤ 314, by comparing the alpha decay half lives and the corresponding spontaneous fission half lives of respective isotopes, may open up new lines in experimental investigations.

## 4. Conclusion

In the present paper we have shown the theoretical predictions on the alpha decay half lives of various isotopes of the element Z = 113, within the Coulomb and proximity potential for the deformed nuclei (CPPMDN). We could successfully reproduce the alpha half lives and decay chains for the experimentally synthesized isotopes $^{278}$113 and $^{282}$113. Hence an extensive study has been done for predicting the alpha decay half lives and decay chains of all the other isotopes in this region. Through our study we understood that isotopes of Z = 113 within the range 278 ≤ A ≤ 286 will survive fission and can be synthesized and detected in laboratories. We have predicted 6α chains from $^{279}$113, 3α chains from $^{280,281,283}$113, 2α chain from $^{284}$113, 1α chain from $^{285}$113 and 4α chains from $^{286}$113. We hope that these predictions will be a guideline for the future experimental investigations.

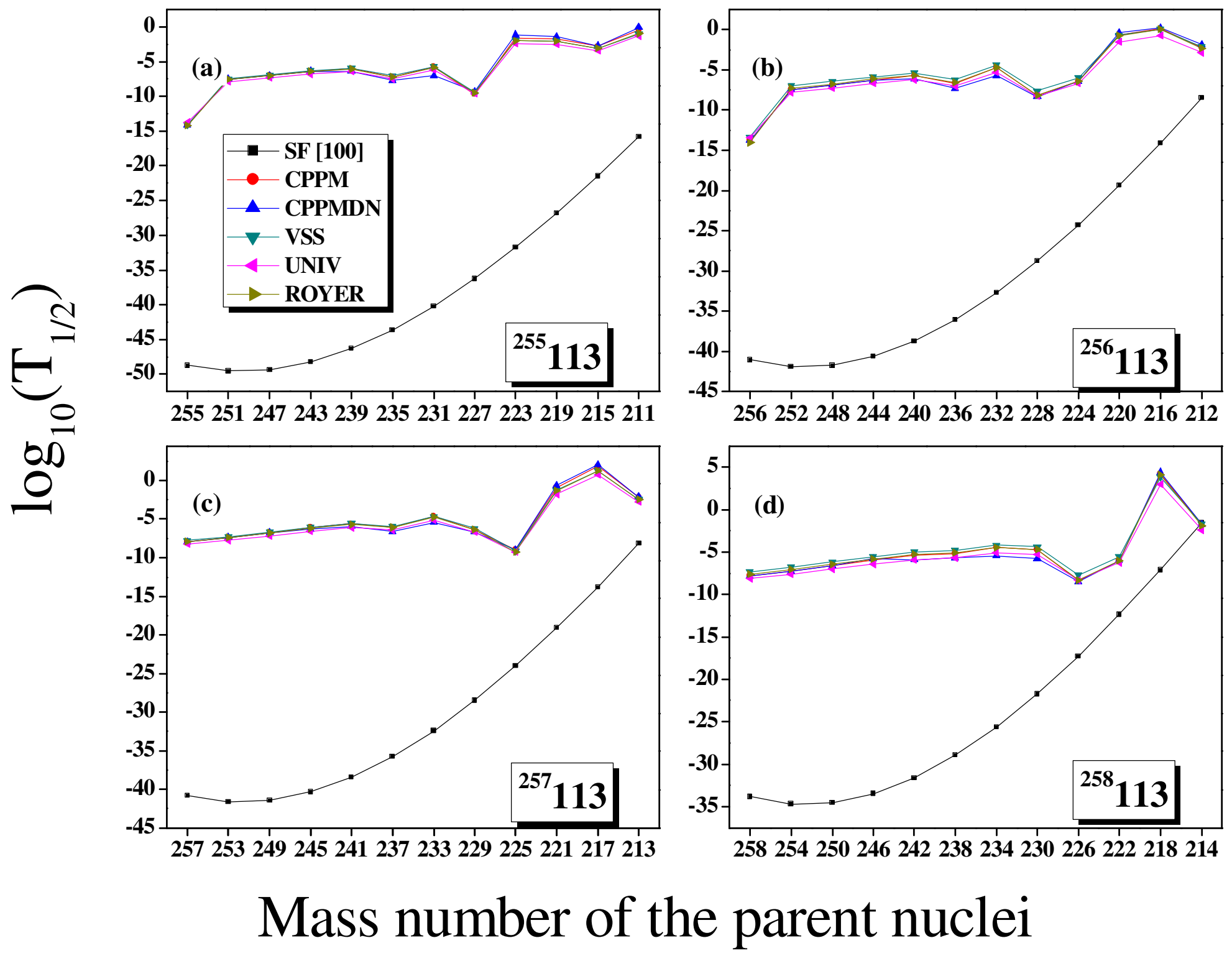

Fig 1: The comparison of the calculated alpha decay half-lives with the spontaneous fission half-lives for the isotopes $^{255\text{-}258}113$.

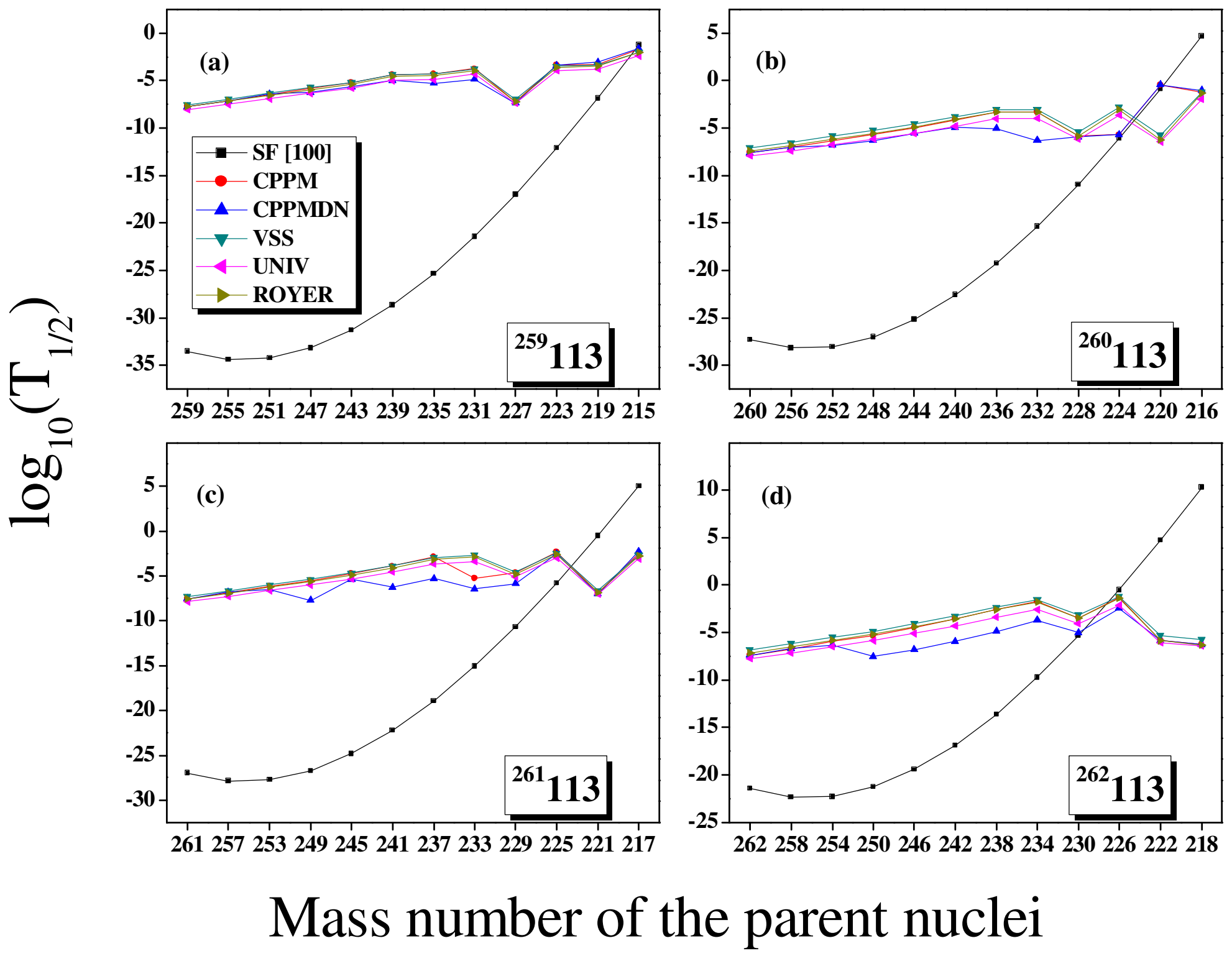

Fig 2: The comparison of the calculated alpha decay half-lives with the spontaneous fission half-lives for the isotopes $^{259-262}113$.

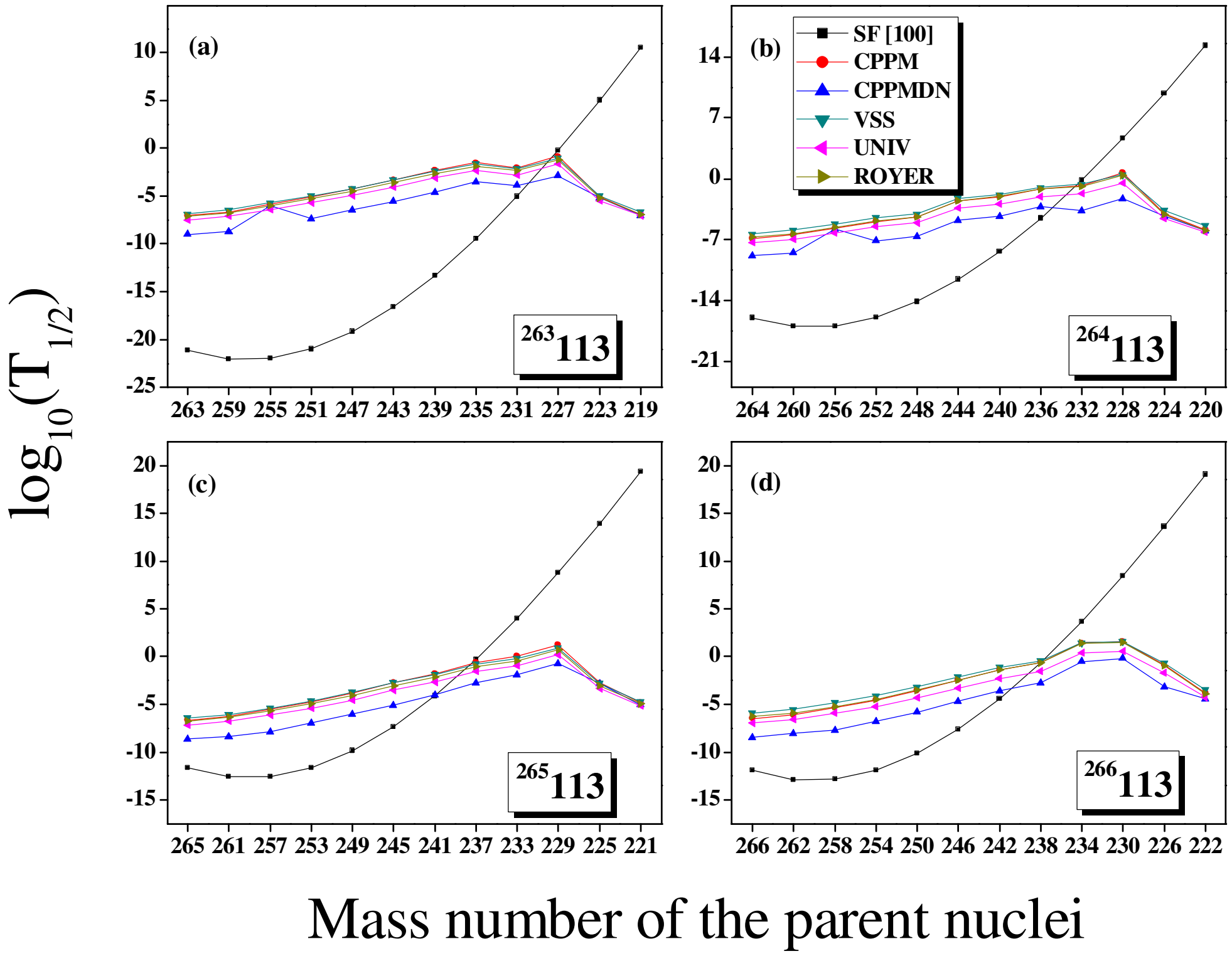

Fig 3: The comparison of the calculated alpha decay half-lives with the spontaneous fission half-lives for the isotopes $^{263-266}113$.

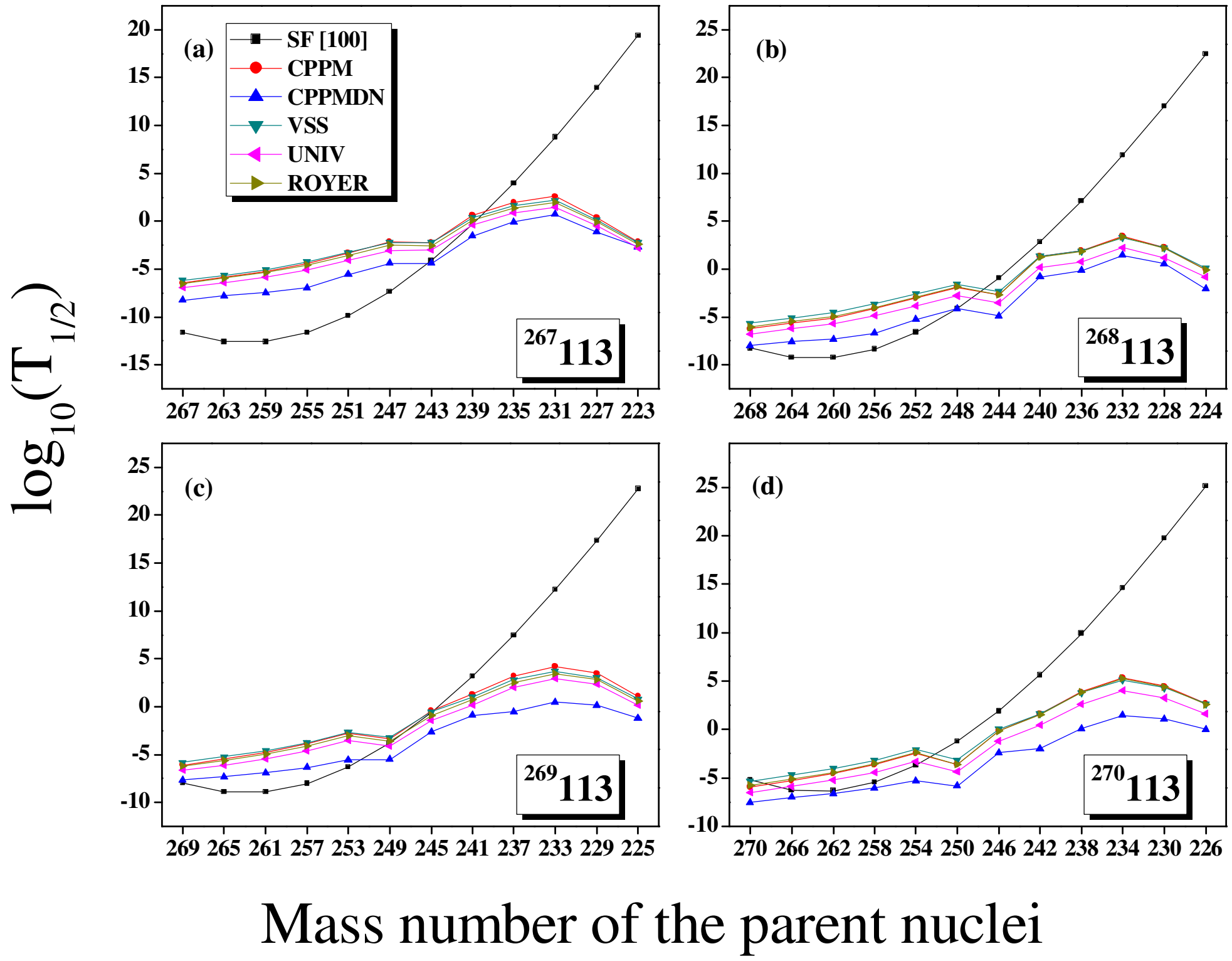

Fig 4: The comparison of the calculated alpha decay half-lives with the spontaneous fission half-lives for the isotopes $^{267\text{-}270}113$.

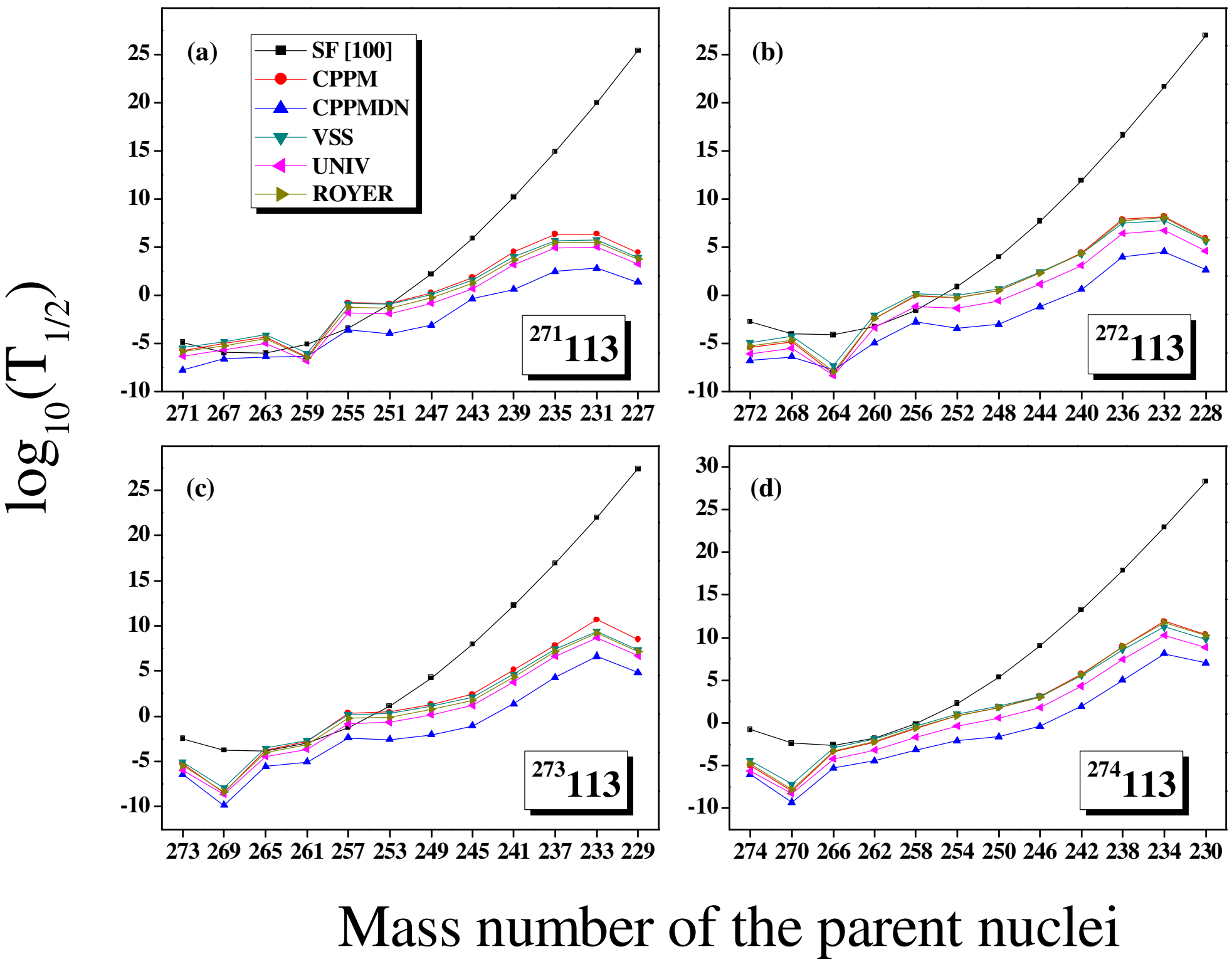

Fig 5: The comparison of the calculated alpha decay half-lives with the spontaneous fission half-lives for the isotopes $^{271\text{-}274}113$.

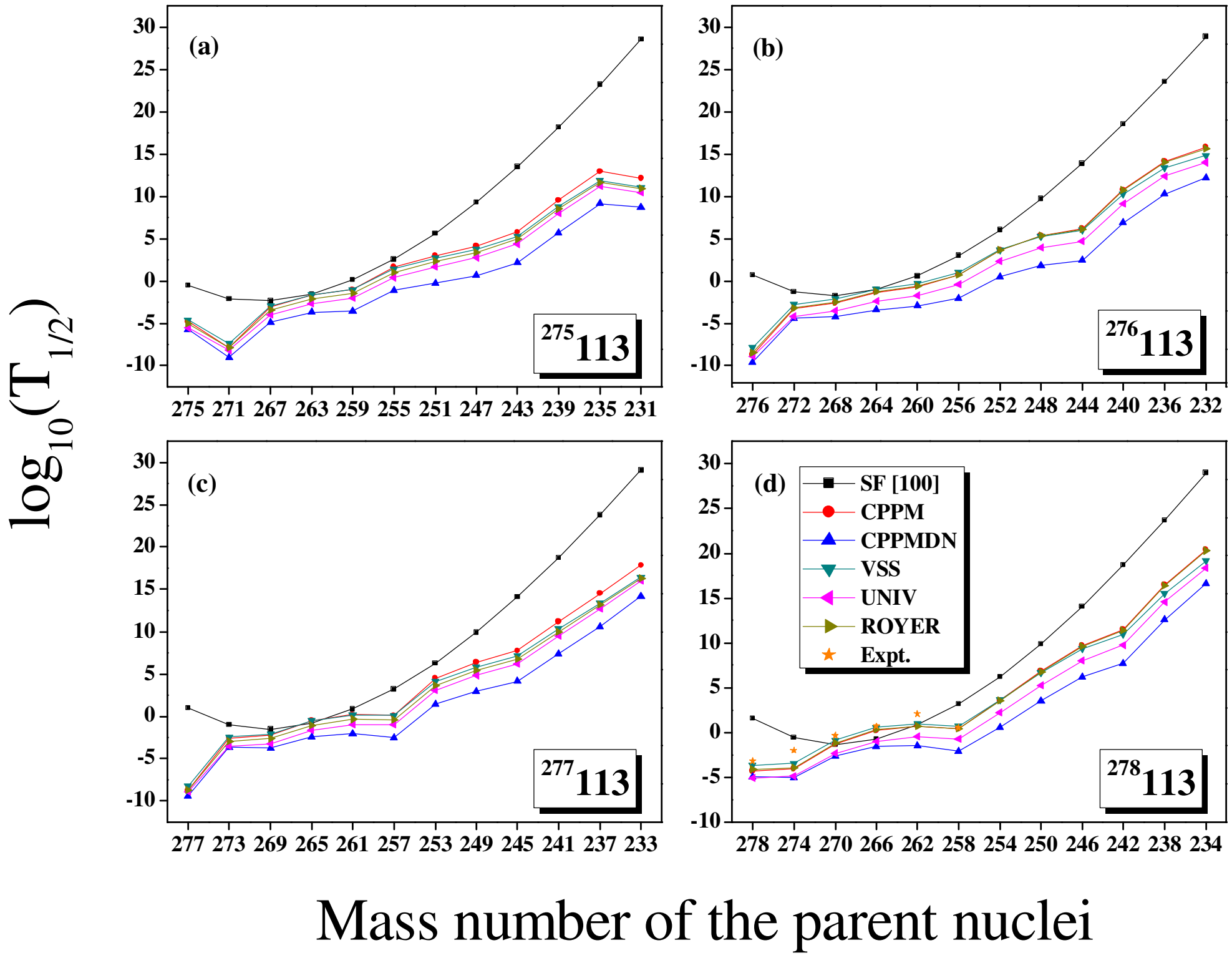

Fig 6: The comparison of the calculated alpha decay half-lives with the spontaneous fission half-lives for the isotopes $^{275\text{-}278}113$.

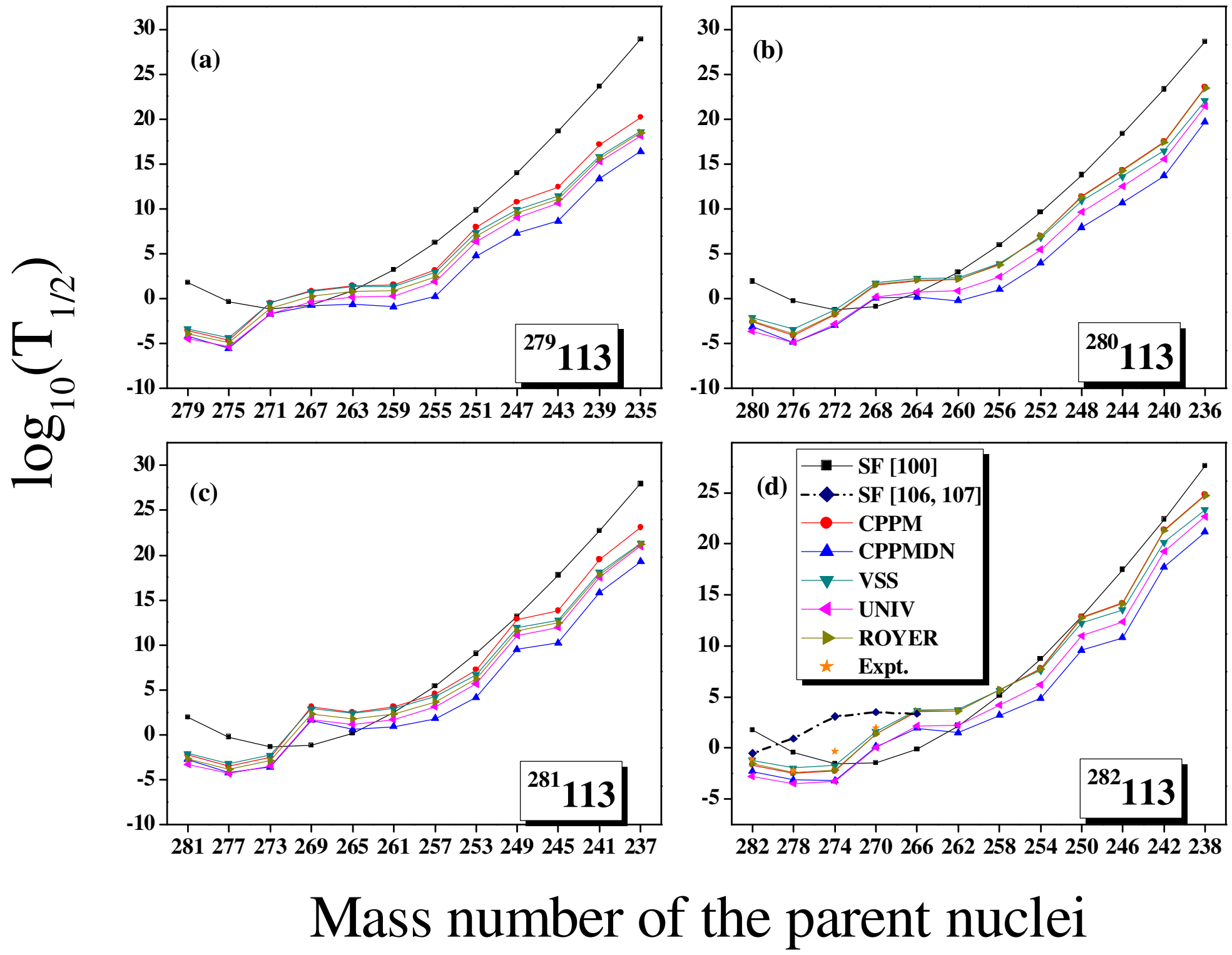

Fig 7: The comparison of the calculated alpha decay half-lives with the spontaneous fission half-lives for the isotopes $^{279-282}113$.

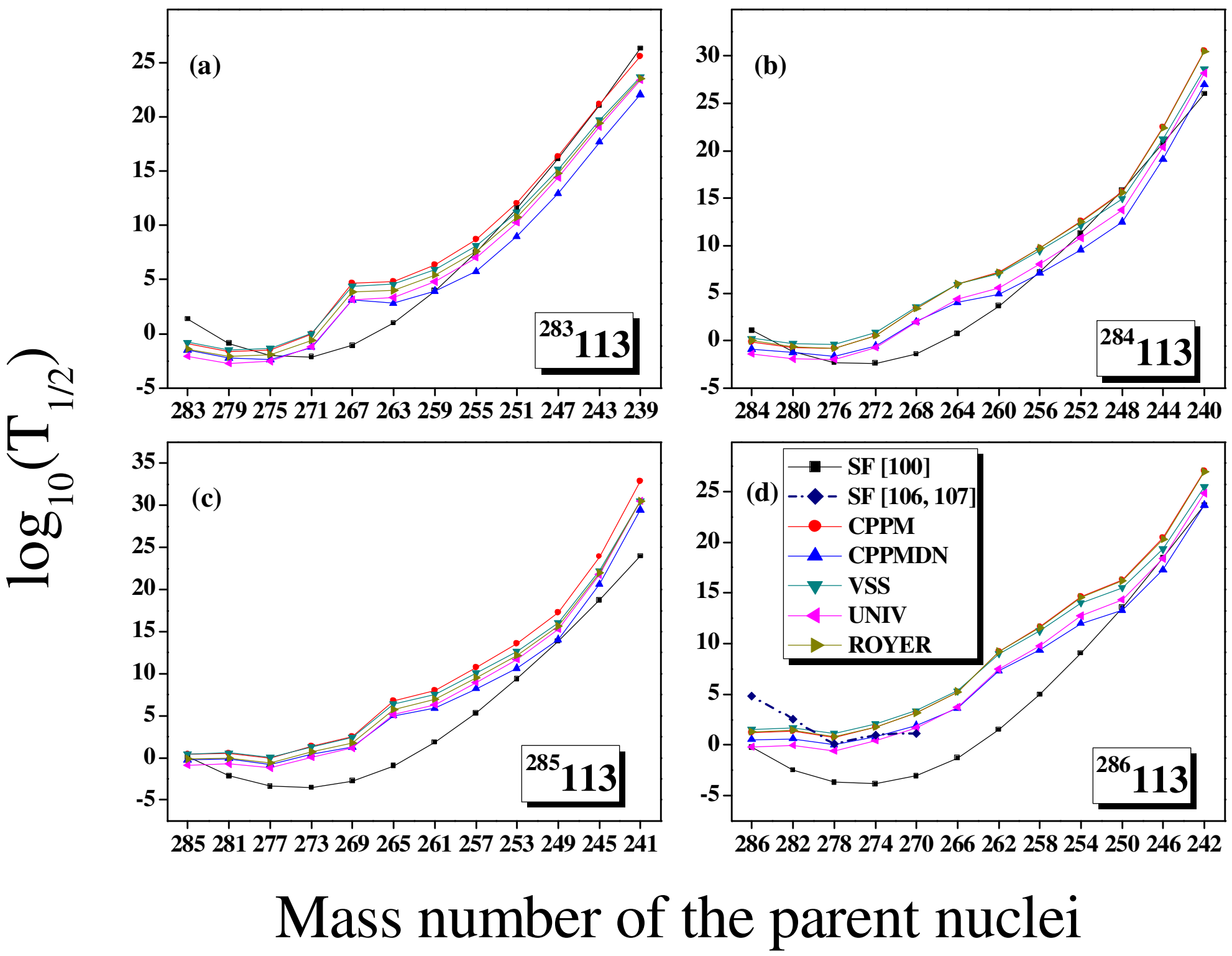

Fig 8: The comparison of the calculated alpha decay half-lives with the spontaneous fission half-lives for the isotopes $^{283-286}$113.

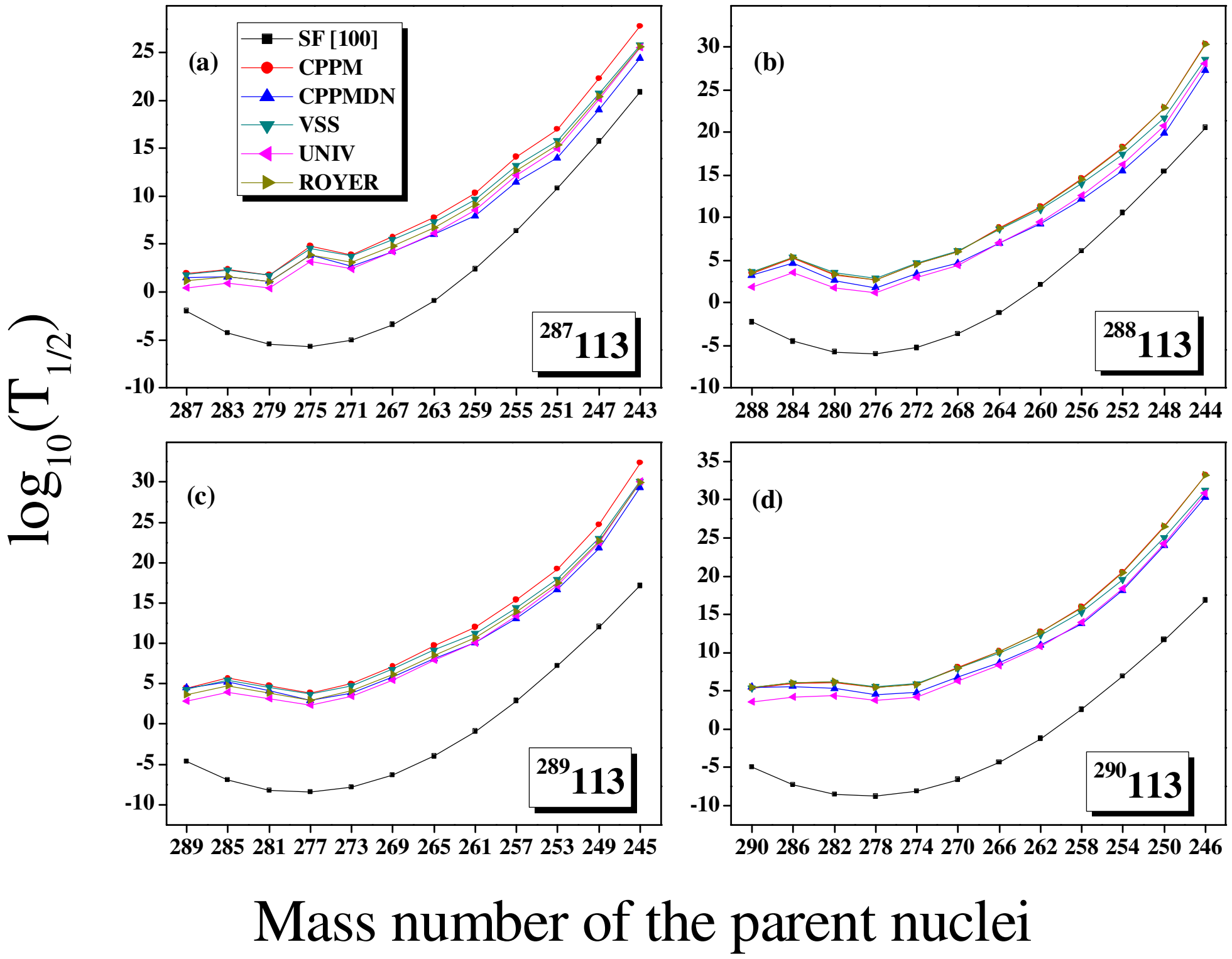

Fig 9: The comparison of the calculated alpha decay half-lives with the spontaneous fission half-lives for the isotopes $^{287\text{-}290}113$.

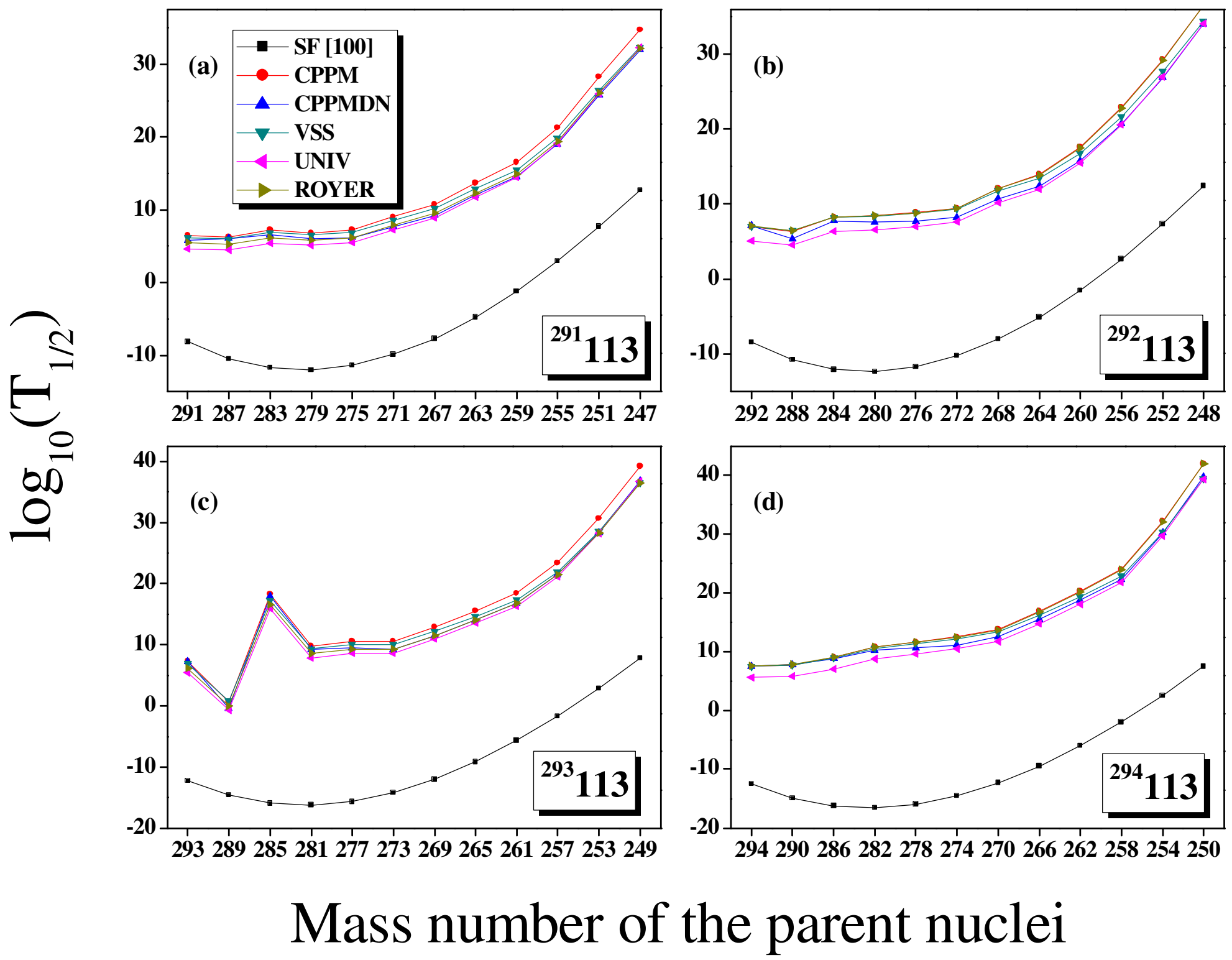

Fig 10: The comparison of the calculated alpha decay half-lives with the spontaneous fission half-lives for the isotopes $^{291\text{-}294}113$.

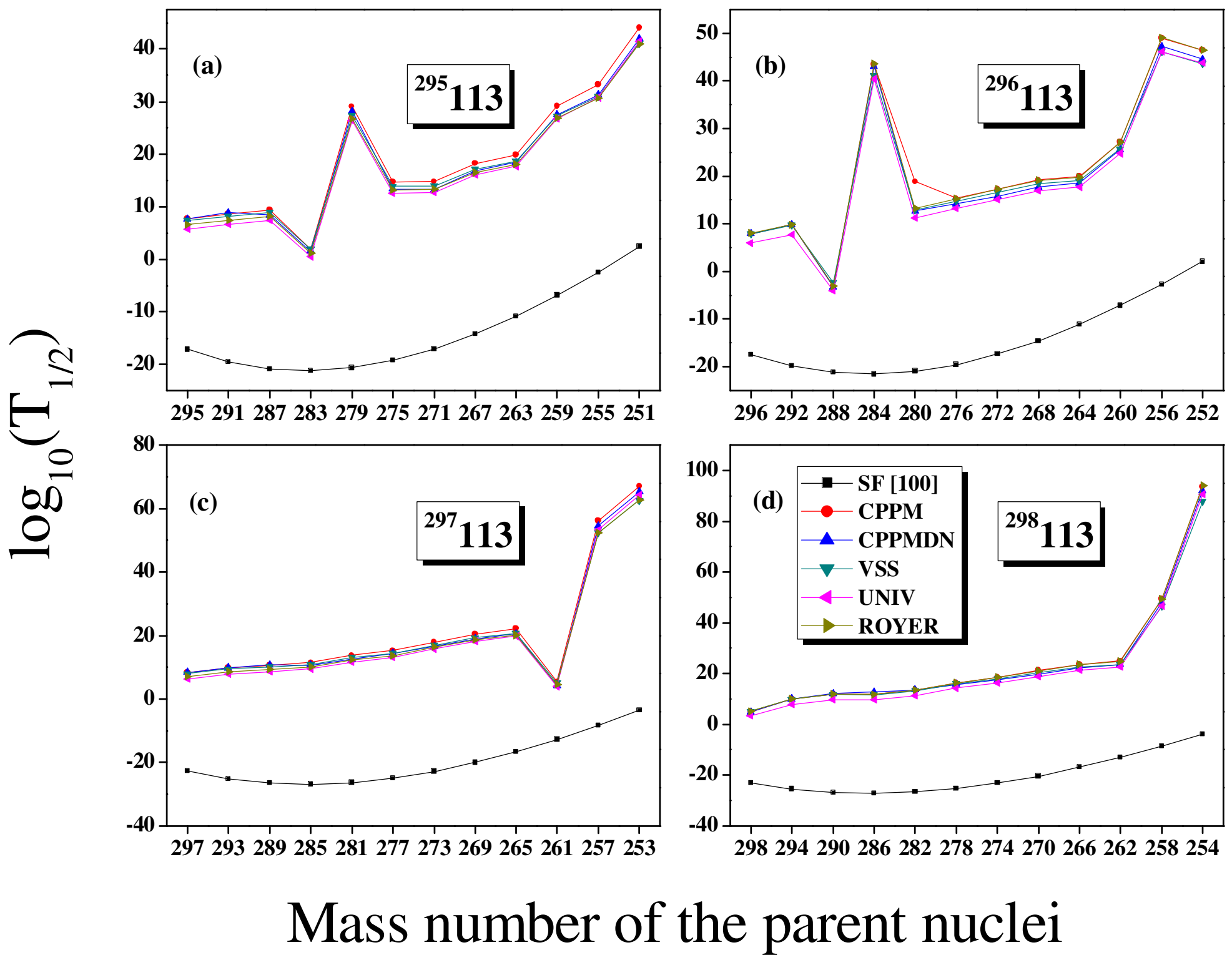

Fig 11: The comparison of the calculated alpha decay half-lives with the spontaneous fission half-lives for the isotopes $^{294\text{-}298}$113.

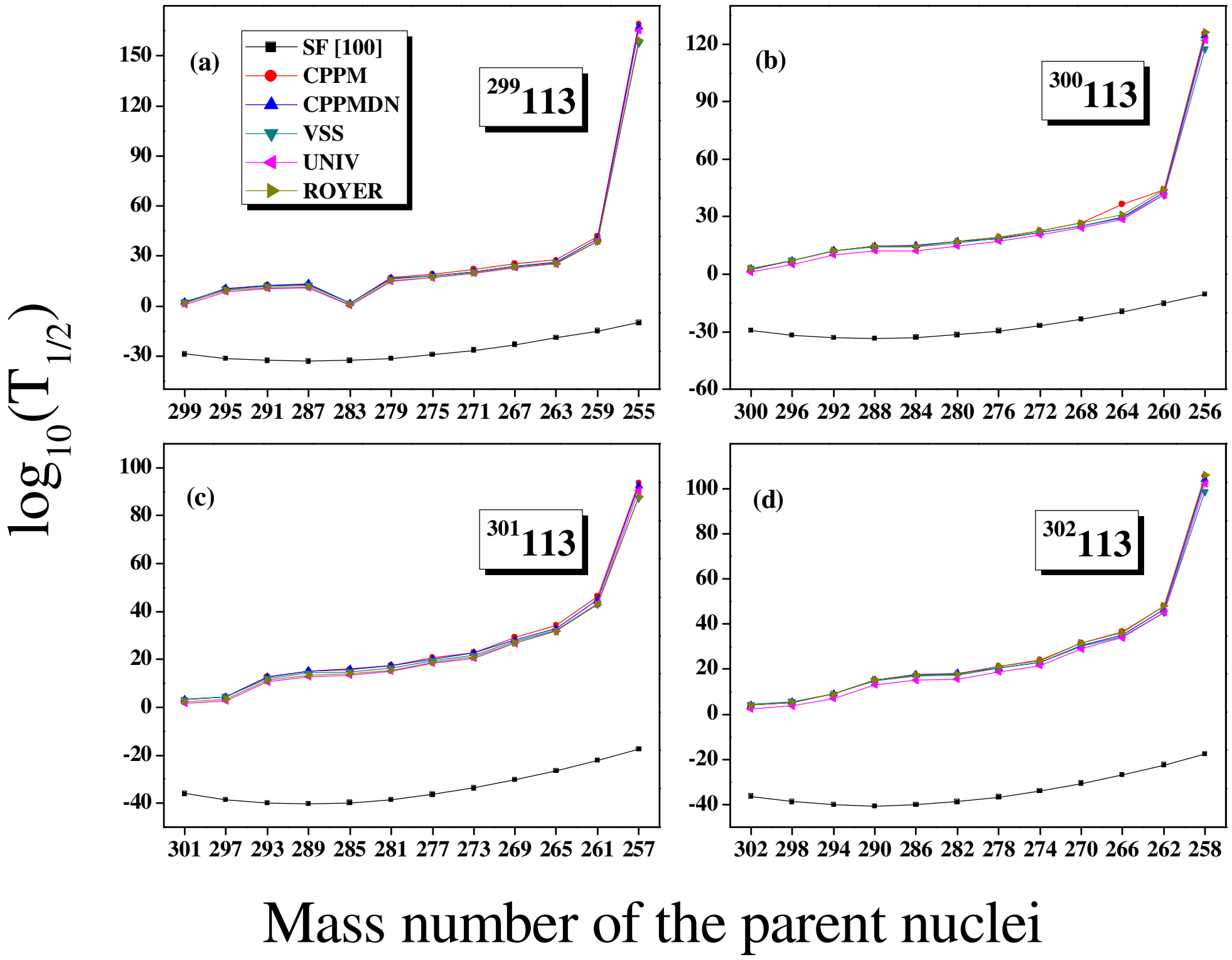

Fig 12: The comparison of the calculated alpha decay half-lives with the spontaneous fission half-lives for the isotopes $^{299\text{-}302}113$.

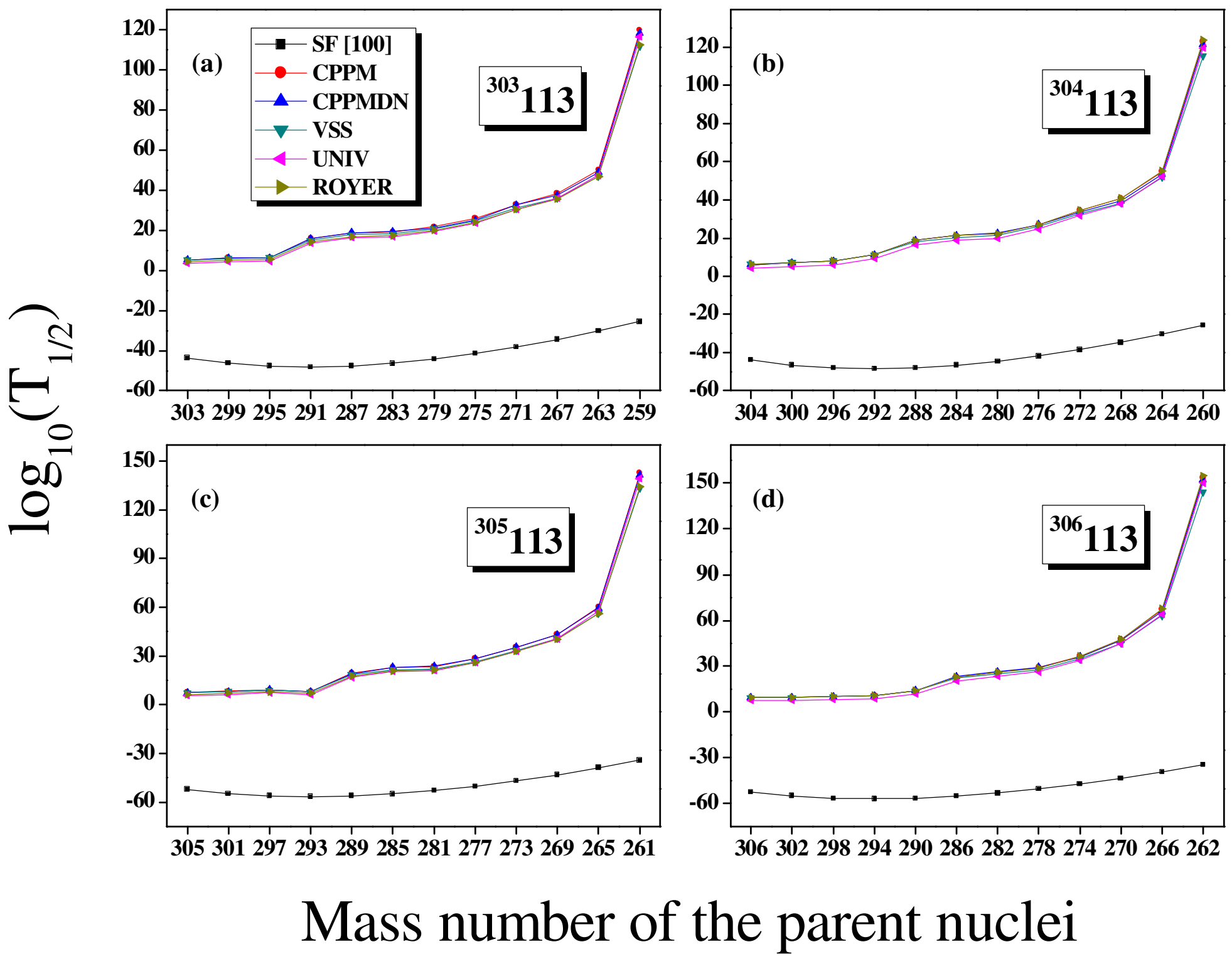

Fig 13: The comparison of the calculated alpha decay half-lives with the spontaneous fission half-lives for the isotopes $^{303\text{-}306}113$.

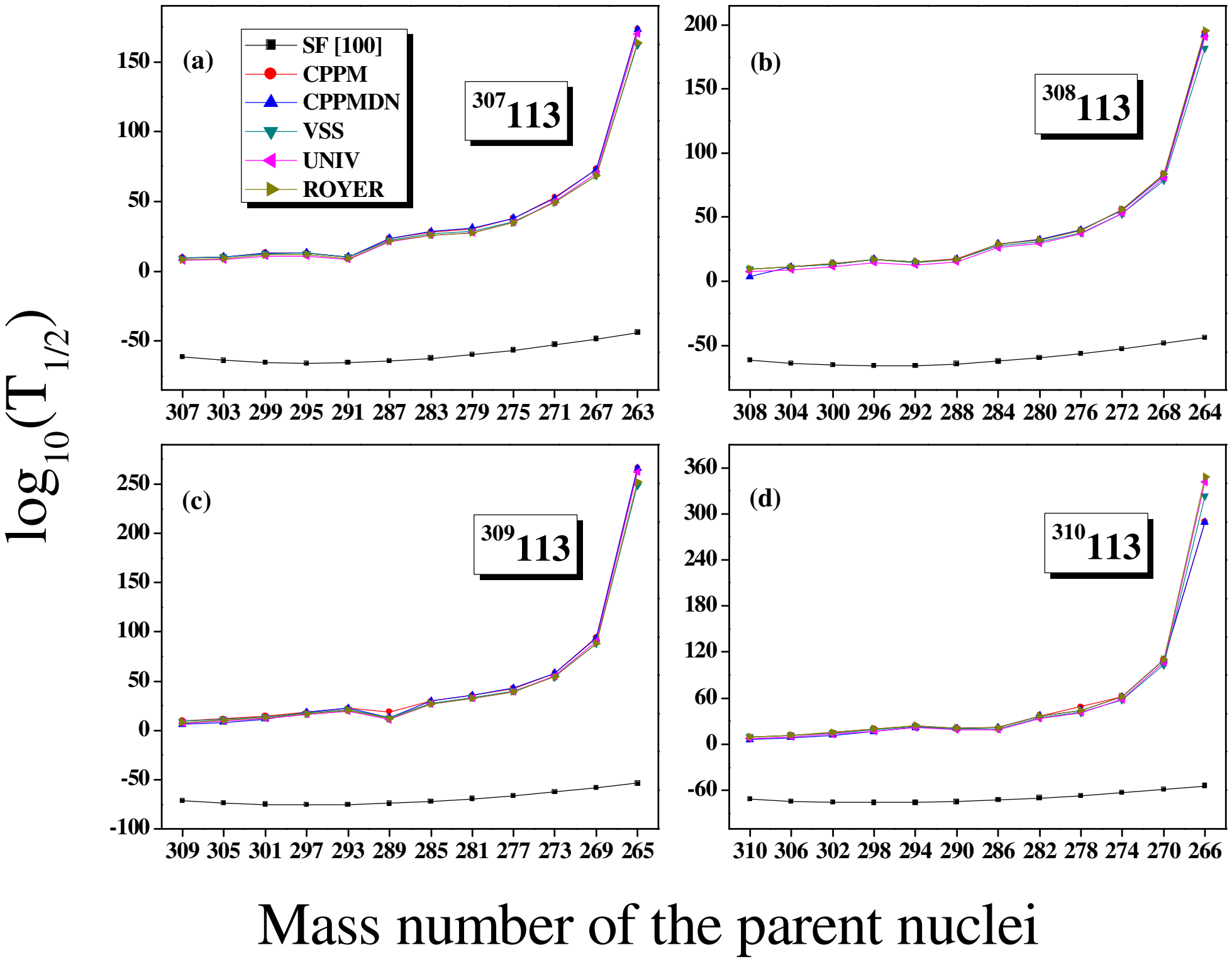

Fig 14: The comparison of the calculated alpha decay half-lives with the spontaneous fission half-lives for the isotopes $^{307-310}113$.

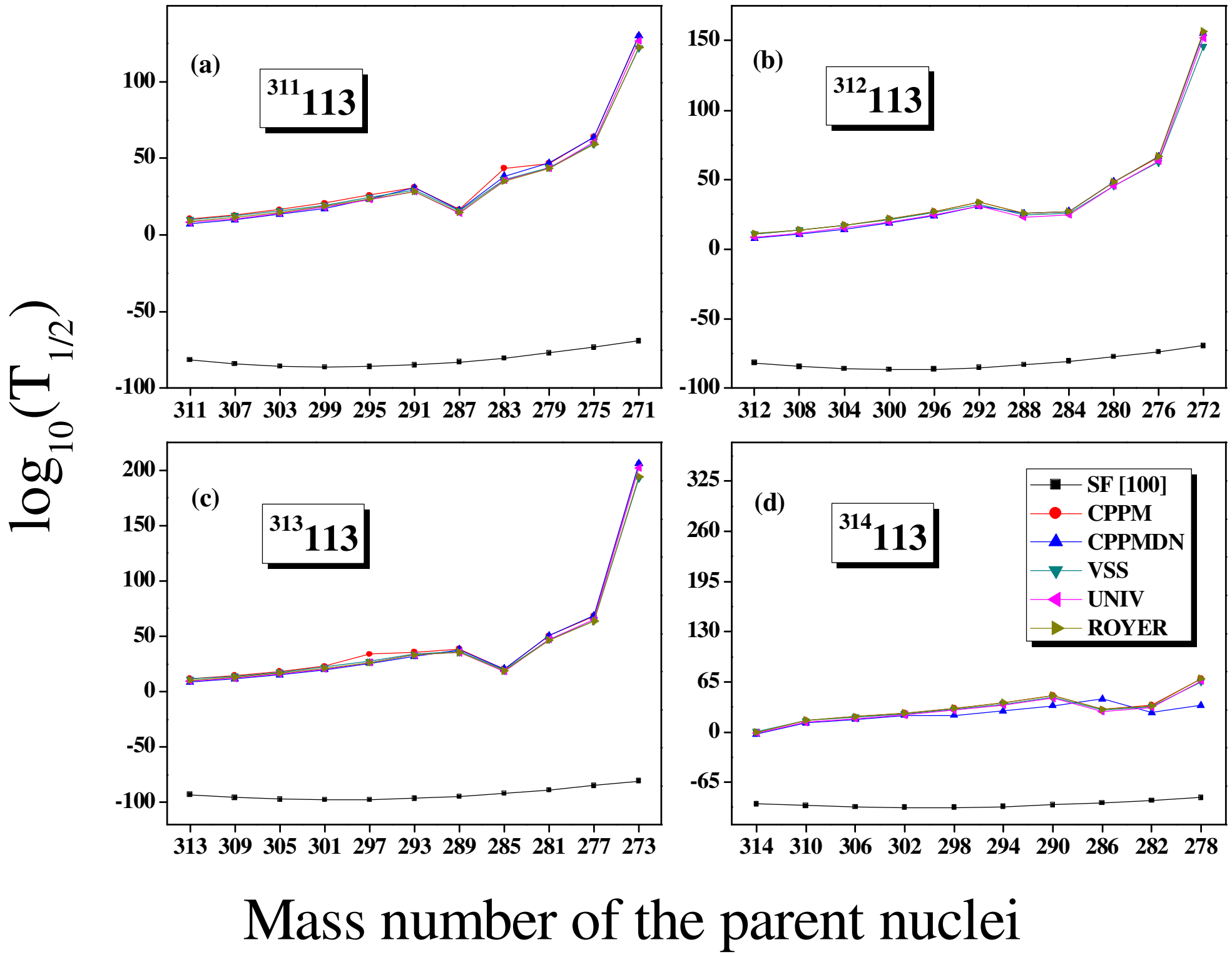

Fig15: The comparison of the calculated alpha decay half-lives with the spontaneous fission half-lives for the isotopes $^{311\text{-}314}113$.

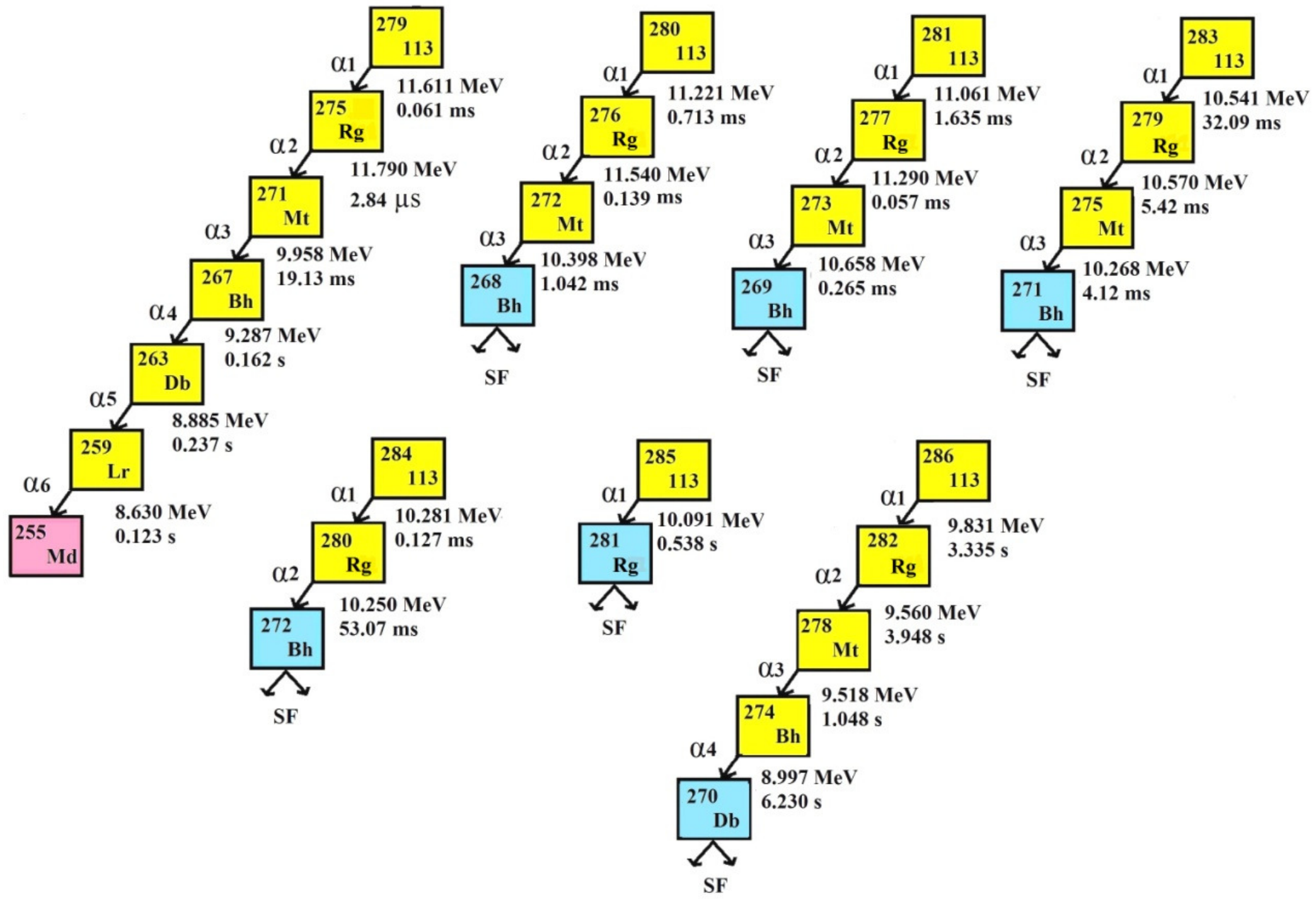

Fig 16: Predicted decay chains for $^{279\text{-}281}113$ and $^{283\text{-}285}113$ isotopes within CPPMDN. The calculated Q values and decay times are shown.

Table I: The comparison of the calculated alpha decay half lives with the spontaneous fission half lives for the isotopes $^{278,282}113$ and its decay products. $T_{SF}^{av}$ is calculated using Ref [100].

| Parent Nuclei | $Q_\alpha$ (Exp) MeV | $T_{SF}^{av}$ (S) | $T_{1/2}^{\alpha}$ | | | | | | Mode of Decay |
|---|---|---|---|---|---|---|---|---|---|
| | | | Expt | CPPMDN | CPPM | VSS | UNIV | Royer | |
| $^{278}113$ | 11.82±0.06 | $3.831\text{x}10^{1}$ | 0.667ms | $0.02^{-0.01}_{+0.01}$ ms | $0.08^{-0.03}_{+0.02}$ ms | $0.31^{-0.11}_{+0.03}$ ms | $0.01^{-0.004}_{+0.003}$ ms | $0.12^{-0.05}_{+0.03}$ ms | α1 |
| $^{274}Rg$ | 10.65±0.06 | $2.853\text{x}10^{-1}$ | 9.97ms | $1.94^{-0.89}_{+0.61}$ ms | $17.39^{-8.01}_{+5.45}$ ms | $46.33^{-19.49}_{+13.63}$ ms | $1.50^{-0.61}_{+0.43}$ ms | $22.87^{-10.44}_{+7.12}$ ms | α2 |
| $^{270}Mt$ | 10.26±0.07 | $4.686\text{x}10^{-2}$ | 444ms | $19.75^{-11.40}_{+18.49}$ ms | $43.46^{-25.31}_{+15.87}$ ms | $111.70^{-59.20}_{+38.42}$ ms | $3.69^{-1.88}_{+1.24}$ ms | $53.63^{-30.98}_{+19.48}$ ms | α3 |
| $^{266}Bh$ | 9.39±0.06 | $1.811\text{x}10^{-1}$ | 5.26s | $0.06^{-0.03}_{+0.02}$ s | $3.37^{-1.86}_{+1.34}$ s | $6.44^{-3.25}_{+2.42}$ s | $0.20^{-0.09}_{+0.08}$ s | $3.83^{-2.10}_{+1.52}$ s | α4 |
| $^{262}Db$ | 8.63±0.06 | $6.941\text{x}10^{0}$ | 126s | $1.34^{-0.89}_{+0.51}$ s | $201.50^{-127.28}_{+77.40}$ s | $290.80^{-167.30}_{+105.40}$ s | $9.28^{-5.26}_{+3.33}$ s | $213.20^{-134.20}_{+81.70}$ s | α5 |
| $^{258}Lr$ | 8.66±0.06 | $1.603\text{x}10^{3}$ | 3.78s | $0.07^{-0.05}_{+0.03}$ s | $26.15^{-16.03}_{+9.86}$ s | $43.73^{-24.39}_{+15.52}$ s | $1.57^{-0.86}_{+0.55}$ s | $26.27^{-16.03}_{+9.87}$ s | α6 |
| $^{254}Md$ | - | - | - | - | - | - | - | - | *EC |
| | | | | | | | | | |
| $^{282}113$ | 10.63±0.08 | $3.023\text{x}10^{-1\#}$ | $73^{+134}_{-29}$ ms | $18.4^{-12.31}_{+7.31}$ ms | $80.9^{-54.6}_{+32.3}$ ms | $214.5^{-131.8}_{+81}$ ms | $5.4^{-1.745}_{+3.738}$ ms | $110.0^{-73.7}_{+43.71}$ ms | α1 |
| $^{278}Rg$ | 10.69±0.08 | $8.388\text{x}10^{0\#}$ | $4.2^{+7.5}_{-1.7}$ ms | $2.6^{-1.03}_{+1.72}$ ms | $11.6^{-4.5}_{+7.6}$ ms | $36.7^{-13.6}_{+21.8}$ ms | $1.0^{-0.369}_{+0.585}$ ms | $15.1^{-5.9}_{+9.8}$ ms | α2 |
| $^{274}Mt$ | 10.0±1.10 | $1.221\text{x}10^{3\#}$ | $0.44^{+0.81}_{-0.17}$ s | $.023^{-75.24}_{+0.023}$ s | $0.21^{-700.9}_{+0.210}$ s | $0.55^{-10.44}_{+0.55}$ s | $0.015^{-23.94}_{+0.149}$ ms | $0.25^{-817.35}_{+0.252}$ s | α3 |
| $^{270}Bh$ | 8.93±0.08 | $3.329\text{x}10^{3\#}$ | $61^{+292}_{-28}$ s | $5.52^{-2.56}_{+4.84}$ s | $94.26^{-43.8}_{+83.2}$ s | $163.70^{-72.1}_{+131.1}$ s | $4.15^{-1.806}_{+3.258}$ s | $104.00^{-98.4}_{+91.6}$ s | α4 |
| $^{266}Db$ | 8.265* | $2.121\text{x}10^{3\#}$ | - | 86.14s | 3694.00s | 4970.00s | 131.50 | 3798.00s | SF |

*Q value computed using experimental mass excess [85]

# $T_{SF}^{av}$ calculated using Ref [106, 107]

Table II: Predictions on the mode of decay of $^{279-281}113$ and $^{283,284}113$ superheavy nuclei and their decay products by comparing the alpha half lives and the corresponding spontaneous fission half lives. $T_{SF}^{av}$ is calculated using Ref [100].

| Parent Nuclei | $Q_\alpha$ (Cal) MeV | $T_{SF}^{av}$ (S) | $T_{1/2}$(s) | | | | | Mode of Decay |
|---|---|---|---|---|---|---|---|---|
| | | | CPPMDN | CPPM | VSS | UNIV | Royer | |
| $^{279}113$ | 11.611 | $6.676 \times 10^{1}$ | $6.074 \times 10^{-5}$ | $2.591 \times 10^{-4}$ | $4.118 \times 10^{-4}$ | $3.229 \times 10^{-5}$ | $1.185 \times 10^{-4}$ | α1 |
| $^{275}$Rg | 11.790 | $4.715 \times 10^{-1}$ | $2.844 \times 10^{-6}$ | $2.257 \times 10^{-5}$ | $4.483 \times 10^{-5}$ | $4.084 \times 10^{-6}$ | $1.245 \times 10^{-5}$ | α2 |
| $^{271}$Mt | 9.958 | $6.248 \times 10^{-2}$ | $1.913 \times 10^{-2}$ | $3.251 \times 10^{-1}$ | $3.282 \times 10^{-1}$ | $2.168 \times 10^{-2}$ | $9.354 \times 10^{-2}$ | α3 |
| $^{267}$Bh | 9.287 | $1.957 \times 10^{-1}$ | $1.619 \times 10^{-1}$ | $7.180 \times 10^{0}$ | $5.931 \times 10^{0}$ | $3.892 \times 10^{-1}$ | $1.712 \times 10^{0}$ | α4 |
| $^{263}$Db | 8.885 | $6.947 \times 10^{0}$ | $2.369 \times 10^{-1}$ | $2.664 \times 10^{1}$ | $2.013 \times 10^{1}$ | $1.395 \times 10^{0}$ | $5.894 \times 10^{0}$ | α5 |
| $^{259}$Lr | 8.630 | $1.540 \times 10^{3}$ | $1.228 \times 10^{-1}$ | $3.357 \times 10^{1}$ | $2.484 \times 10^{1}$ | $1.873 \times 10^{0}$ | $7.395 \times 10^{0}$ | α6 |
| $^{255}$Md | 7.952 | $1.641 \times 10^{6}$ | $1.751 \times 10^{0}$ | $1.532 \times 10^{3}$ | $8.534 \times 10^{2}$ | $6.863 \times 10^{1}$ | $2.688 \times 10^{2}$ | *EC |
| | | | | | | | | |
| $^{280}113$ | 11.221 | $7.832 \times 10^{1}$ | $7.131 \times 10^{-4}$ | $2.290 \times 10^{-3}$ | $7.295 \times 10^{-3}$ | $2.273 \times 10^{-4}$ | $3.211 \times 10^{-3}$ | α1 |
| $^{276}$Rg | 11.540 | $5.148 \times 10^{-1}$ | $1.386 \times 10^{-5}$ | $8.296 \times 10^{-5}$ | $3.513 \times 10^{-4}$ | $1.311 \times 10^{-5}$ | $1.129 \times 10^{-4}$ | α2 |
| $^{272}$Mt | 10.398 | $5.440 \times 10^{-2}$ | $1.042 \times 10^{-3}$ | $1.651 \times 10^{-2}$ | $4.879 \times 10^{-2}$ | $1.539 \times 10^{-3}$ | $2.031 \times 10^{-2}$ | α3 |
| $^{268}$Bh | 9.077 | $1.295 \times 10^{-1}$ | $1.057 \times 10^{0}$ | $3.243 \times 10^{1}$ | $5.676 \times 10^{1}$ | $1.584 \times 10^{0}$ | $3.631 \times 10^{1}$ | SF |
| | | | | | | | | |
| $^{281}113$ | 11.061 | $8.988 \times 10^{1}$ | $1.635 \times 10^{-3}$ | $5.742 \times 10^{-3}$ | $8.066 \times 10^{-3}$ | $5.120 \times 10^{-4}$ | $2.154 \times 10^{-3}$ | α1 |
| $^{277}$Rg | 11.290 | $5.580 \times 10^{-1}$ | $5.684 \times 10^{-5}$ | $3.306 \times 10^{-4}$ | $5.942 \times 10^{-4}$ | $4.401 \times 10^{-5}$ | $1.532 \times 10^{-4}$ | α2 |
| $^{273}$Mt | 10.658 | $4.632 \times 10^{-2}$ | $2.646 \times 10^{-4}$ | $3.106 \times 10^{-3}$ | $4.891 \times 10^{-3}$ | $3.456 \times 10^{-4}$ | $1.251 \times 10^{-3}$ | α3 |
| $^{269}$Bh | 8.617 | $6.328 \times 10^{-2}$ | $3.963 \times 10^{1}$ | $1.221 \times 10^{3}$ | $7.805 \times 10^{2}$ | $4.380 \times 10^{1}$ | $2.132 \times 10^{2}$ | SF |
| | | | | | | | | |
| $^{283}113$ | 10.541 | $2.348 \times 10^{1}$ | $3.209 \times 10^{-2}$ | $1.383 \times 10^{-1}$ | $1.660 \times 10^{-1}$ | $8.709 \times 10^{-3}$ | $4.117 \times 10^{-2}$ | α1 |
| $^{279}$Rg | 10.570 | $1.351 \times 10^{-1}$ | $5.420 \times 10^{-3}$ | $2.401 \times 10^{-2}$ | $3.372 \times 10^{-2}$ | $1.956 \times 10^{-3}$ | $8.131 \times 10^{-3}$ | α2 |
| $^{275}$Mt | 10.268 | $9.492 \times 10^{-3}$ | $4.107 \times 10^{-3}$ | $3.420 \times 10^{-2}$ | $4.832 \times 10^{-2}$ | $2.910 \times 10^{-3}$ | $1.149 \times 10^{-2}$ | α3 |
| $^{271}$Bh | 9.537 | $7.582 \times 10^{-3}$ | $5.983 \times 10^{-2}$ | $9.673 \times 10^{-1}$ | $1.097 \times 10^{0}$ | $6.409 \times 10^{-2}$ | $2.640 \times 10^{-1}$ | SF |
| | | | | | | | | |
| $^{284}113$ | 10.281 | $1.234 \times 10^{1}$ | $1.265 \times 10^{-1}$ | $7.427 \times 10^{-1}$ | $1.802 \times 10^{0}$ | $3.933 \times 10^{-2}$ | $9.883 \times 10^{-1}$ | α1 |
| $^{280}$Rg | 10.250 | $7.076 \times 10^{-2}$ | $5.307 \times 10^{-2}$ | $1.852 \times 10^{-1}$ | $5.107 \times 10^{-1}$ | $1.215 \times 10^{-2}$ | $2.329 \times 10^{-1}$ | α2 |
| $^{276}$Mt | 10.048 | $4.952 \times 10^{-3}$ | $2.425 \times 10^{-2}$ | $1.415 \times 10^{-1}$ | $4.098 \times 10^{-1}$ | $1.035 \times 10^{-2}$ | $1.683 \times 10^{-1}$ | SF |

Table III: Predictions on the mode of decay of $^{285,286}113$ superheavy nuclei and their decay products by comparing the alpha half lives and the corresponding spontaneous fission half lives. $T_{SF}^{av}$ is calculated using Ref [100].

| Parent Nuclei | $Q_\alpha$ (Cal) MeV | $T_{SF}^{av}$ (S) | $T_{1/2}$(s) | | | | | Mode of Decay |
|---|---|---|---|---|---|---|---|---|
| | | | CPPMDN | CPPM | VSS | UNIV | Royer | |
| $^{285}113$ | 10.091 | $1.186\times10^{0}$ | $5.381\times10^{-1}$ | $2.617\times10^{0}$ | $2.740\times10^{0}$ | $1.221\times10^{-1}$ | $6.309\times10^{-1}$ | α1 |
| $^{281}$Rg | 9.820 | $6.390\times10^{-3}$ | $7.702\times10^{-1}$ | $3.408\times10^{0}$ | $3.602\times10^{0}$ | $1.675\times10^{-1}$ | $8.147\times10^{-1}$ | SF |
| $^{286}113$ | 9.831 | $6.259\times10^{4\#}$ | $3.335\times10^{0}$ | $1.577\times10^{1}$ | $3.319\times10^{1}$ | $6.196\times10^{-1}$ | $2.055\times10^{1}$ | α1 |
| $^{282}$Rg | 9.560 | $3.258\times10^{2\#}$ | $3.948\times10^{0}$ | $2.147\times10^{1}$ | $4.538\times10^{1}$ | $8.862\times10^{-1}$ | $2.619\times10^{1}$ | α2 |
| $^{278}$Mt | 9.518 | $1.236\times10^{0\#}$ | $1.048\times10^{0}$ | $5.371\times10^{0}$ | $1.283\times10^{1}$ | $2.736\times10^{-1}$ | $6.210\times10^{0}$ | α3 |
| $^{274}$Bh | 8.977 | $1.018\times10^{1\#}$ | $6.230\times10^{0}$ | $5.656\times10^{1}$ | $1.165\times10^{2}$ | $2.545\times10^{0}$ | $6.120\times10^{1}$ | α4 |
| $^{270}$Db | 8.365 | $1.335\times10^{1\#}$ | $8.365\times10^{1}$ | $1.361\times10^{3}$ | $2.240\times10^{3}$ | $5.106\times10^{1}$ | $1.367\times10^{3}$ | SF |

# $T_{SF}^{av}$ calculated using Ref [106, 107]